\documentclass[10pt,twocolumn,superscriptaddress,longbibliography,prx]{revtex4-2}

\usepackage{amsmath, bm, amsfonts, amssymb}     
\usepackage{graphicx}                       
\usepackage{xfrac}                          
\usepackage{grffile}                        
\usepackage{color,colortbl}
\usepackage{xcolor}

\newcommand{\gdot}[0] {\dot{\gamma}}

\begin{document}


\title{Discontinuous shear thickening in biological tissue rheology} 

\author{Michael J. Hertaeg}
\affiliation{%
Department of Physics, Durham University, Science Laboratories, South Road, Durham DH1 3LE, UK
}
\author{Suzanne M. Fielding}%
\affiliation{%
Department of Physics, Durham University, Science Laboratories, South Road, Durham DH1 3LE, UK
}%

\author{Dapeng Bi}
\affiliation{%
Department of Physics, Northeastern University, MA 02115, USA
}%

\date{\today}
\begin{abstract}
During embryonic morphogenesis, tissues undergo dramatic deformations in order to form functional organs. Similarly, in adult animals, living cells and tissues are continually subjected to forces and deformations. Therefore, the success of embryonic development and the proper maintenance of physiological functions rely on the ability of cells to withstand mechanical stresses as well as their ability to flow in a collective manner. During these events, mechanical perturbations can originate from active processes at the single cell level, competing with external stresses exerted by surrounding tissues and organs. However, the study of tissue mechanics has been somewhat limited to either the response to external forces or to intrinsic ones. In this work, we use an active vertex model of a 2D confluent tissue to study the interplay of external deformations that are applied globally to a tissue with internal active stresses that arise locally at the cellular level due to cell motility. We elucidate in particular the way in which this interplay between globally external and locally internal active driving determines the emergent mechanical properties of the tissue as a whole. For a tissue in the vicinity of a solid-fluid jamming/unjamming transition, we uncover a host of fascinating rheological phenomena, including yielding, shear thinning, continuous shear thickening (CST) and discontinuous shear thickening (DST). These model predictions provide a framework for understanding the recently observed nonlinear rheological behaviors {\it in vivo}.
\end{abstract}

\maketitle

\section{Introduction}
During embryonic morphogenesis, biological tissues undergo dramatic deformations  in order to form functional organs. The tissues of mature organisms likewise continually suffer stresses and deformations. The success of embryonic development and the maintenance of proper physiological functioning accordingly both depend intimately on a tissue's rheological (deformation and flow) properties~\cite{petridou_review_rheology}.  On short timescales, tissues can withstand mechanical stresses. Over longer timescales they remodel via cell neighbour exchanges (topological T1 transitions)~\cite{Irvine827, Walck-Shannon2013,Bi_SM_2014,Das_neighbor_exchange}, which thus constitute a key rate-limiting step in important processes such as embryo development, wound healing and cancer metastasis. Recent evidence further suggests that dense confluent tissues, which have no gaps between cells, are poised in the vicinity of a transition between a jammed, solid-like state and an unjammed, fluid-like state
~\cite{Park_NMAT_2015,Garcia2015,Oswald2017,malinverno2017endocytic,Atia2018GeometricJamming,Mongera2018,Ilina2020,Mitchel2020,Petridou_percolation}. 

From a fundamental viewpoint, mechanical stresses can either originate \emph{internally} within a biological tissue, via spontaneously active processes intrinsic to the cellular level, such as cell contractility~\cite{noll2017active,streichan_elife_2018}, polarized motility~\cite{Rappel_polarity_review_2017} or mitosis~\cite{firmino_dev_cell_2016,petridou_ncb_2019}; or they can be exerted \emph{externally}, by surrounding tissues and organs~\cite{fernandez2015mechanical, Wang_anisotropy_fly}.
Recent experiments have shown that cell collectives subjected to externally applied stretching~\cite{trepat_fredberg_stretch_nature_2007, Harris_PNAS_stretch, khalilgharibi2019stress, bonfanti_fracture_review_2022} or shear ~\cite{Pruitt_shear_elife} deformations show a strongly non-linear rheological response. Tissues deformed by internally active stresses at the cellular level have likewise been seen to exhibit extreme mechanical phenomena such as fracturing~\cite{Prakash2021}.

Perhaps surprisingly, studies of tissue mechanics to date have largely been confined \emph{either} to the response of tissues to externally imposed stresses \emph{or}, separately, to phenomena arising from internally active processes. Crucially, however, most living tissues exist in a state where \emph{both} forms of driving work together in concert.  For example, during Drosophila embryogenesis, polarized actomyosin contractility at the single cell level interacts with external stresses exerted by neighbouring tissues to cause the tissue to flow plastically in convergent extension~\cite{Wang_anisotropy_fly,lefebvre2023geometric}. During cancer progression, tumour cell collectives constantly experience mechanical stimuli such as compression and shear stresses from the surrounding extracellular matrix (ECM)~\cite{Northcott_fcell_review}. At the same time, tumour cells generate actomyosin contractility at the single cell level. This interplay between external microenvironmental stresses and internal motility has been shown to be central to determining whether a cell cluster is jammed or unjammed~\cite{Ilina2020}.  Recent work on cancer migration also suggests that tumour fluidity depends not only on the single cell invasive potential (akin to our activity) but also on the compressive and shear stresses they experience due to the ECM~\cite{KANG2021103252}.

With these motivations, in this work, we elucidate in particular the way in which the interplay between globally \emph{external} and locally \emph{internal} active driving determines the emergent mechanical properties of the tissue as a whole. Model predictions point towards a framework for understanding the recently observed range of nonlinear rheological behaviors  {\it in vivo}~\cite{Prakash2021,lefebvre2023geometric,Shelton2021_stress_drive_fluidization} and {\it in vitro}~\cite{Harris_PNAS_stretch,Pruitt_shear_elife}. 
For a tissue in the vicinity of a solid-fluid jamming/unjamming transition, we uncover a host of fascinating rheological phenomena, including yielding, shear thinning, continuous shear thickening (CST) and discontinuous shear thickening (DST).

Beyond this context of biological tissues, shear thickening has been the focus of intense recent research in the rheology literature more broadly, due to its widespread occurrence in dense granular materials and suspensions~\cite{Sedes2020,Seto2013,Behringer2019,Bi2011}. Indeed, simulations~\cite{mari2015discontinuous,mari2014shear} and experiments~\cite{bender1996reversible} on dense suspensions show a large discontinuous increase in viscosity with increasing shear rate, attributed to a cross-over between hydrodynamic and frictional interparticle interactions. For shear rates in this transition region, large stress fluctuations are seen, with an intermittent bimodal switching between low and high viscosity branches of the flow curve~\cite{mari2014shear,Sedes2020}. Associated with this shear thickening transition is the formation of bands of different shear stress stacked with layer normals in the vorticity direction~\cite{chacko2018dynamic}.  Our prediction of DST in biological tissues suggests that this phenomenon may be present in a broader class of materials than is evident from this existing rheology literature.

\section{Model}
 The vertex model that we simulate represents the tightly packed cells of a 2D tissue monolayer as a tiling of $n=1\cdots N$ polygons, defined by the positions of the polygon vertices~\cite{Mitchel2020,Duclut2021,Yan2019,Bi2016,Tong2021,Fielding2022}.  The vertices  of any polygon are joined by edges that form the boundaries with the adjoining cells. Each vertex is shared by three cells and each edge by two cells. 
\begin{figure}[b]
\includegraphics[width=0.3\textwidth]{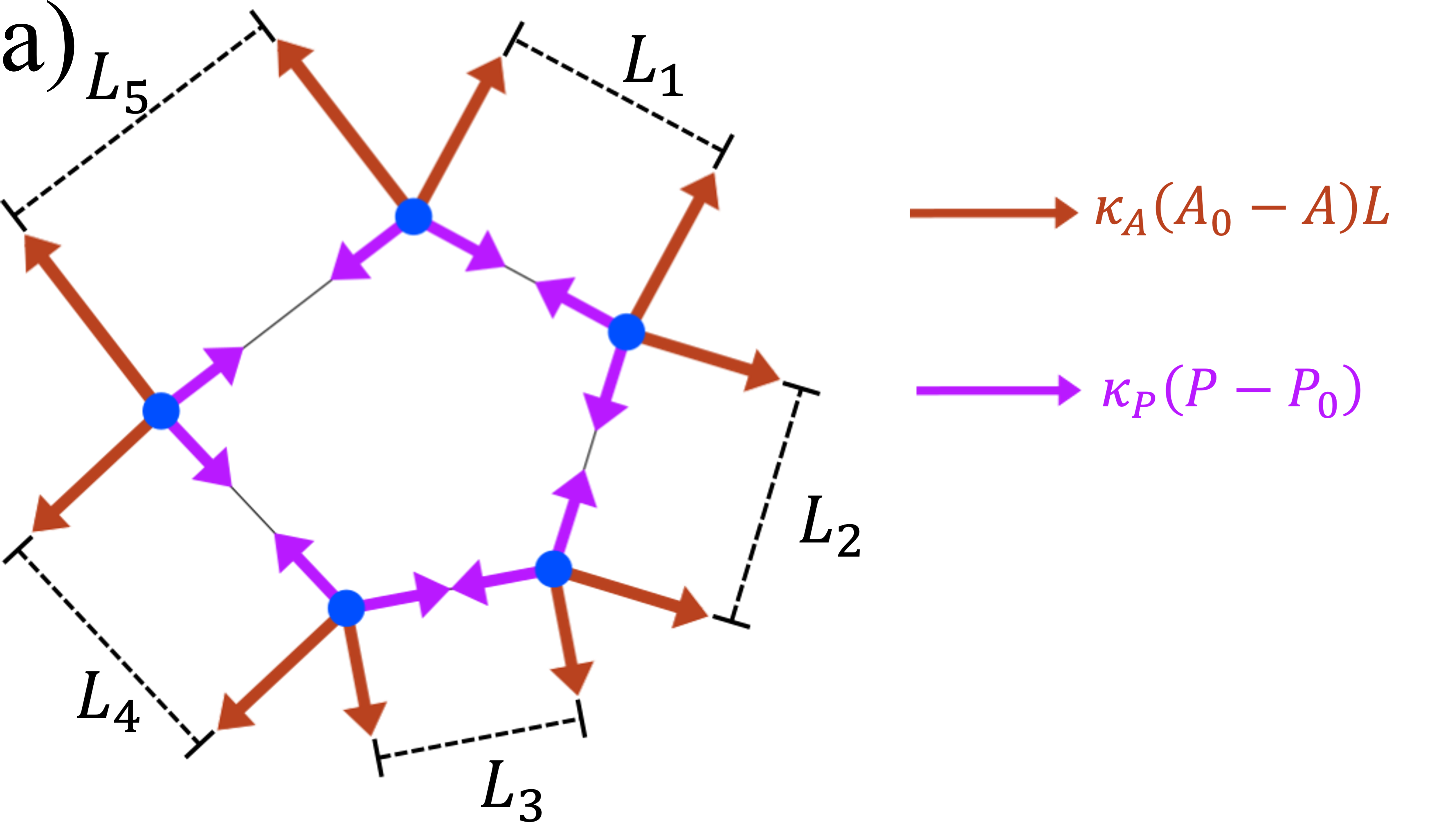}
\includegraphics[width=0.3\textwidth]{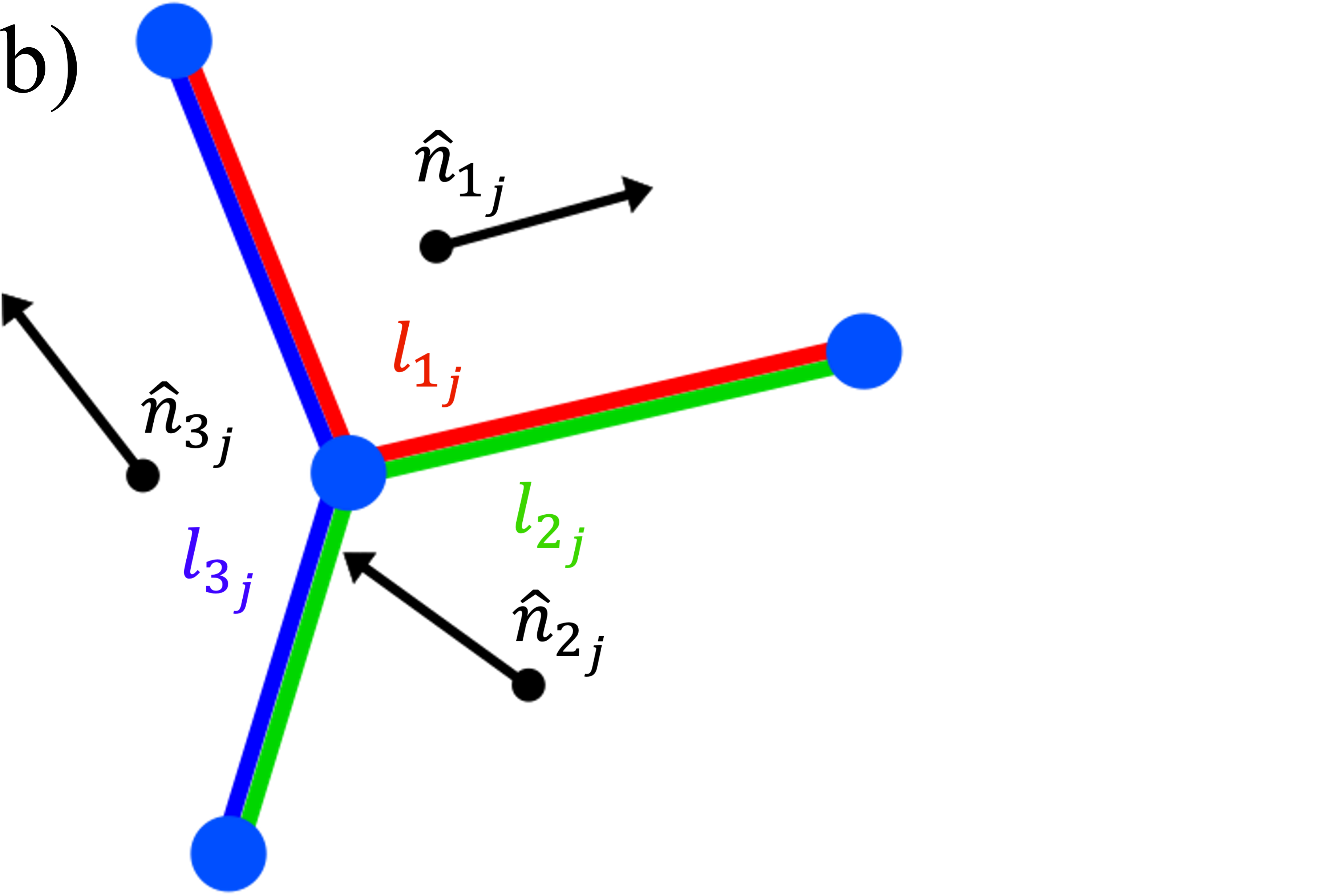}
\caption{{\bf a)} Diagram of elastic forces in the vertex model. Forces tangential to cell edges are proportional to the deviation in cell perimeter and are all of the same magnitude for a single cell. Forces perpendicular to edges are proportional to both the deviation in cell area and the associated edge length. {\bf b)} Diagram of the $j$th vertex in the packing (central blue circle), showing the three edges connecting this vertex to its three neighbouring vertices (other blue circles). The cell polarisation vectors $\hat{n}_{i_j}$ of the three adjoining cells $i_j=1,2,3$ are shown as vectors. The associated lengths $l_{i_j}$ used in the weighted sum to  calculate the polarization vector of the vertex are shown by coloured lines.}
\label{fig:force_diag}
\end{figure}

The elastic energy of the vertex model comprises two contributions. The first is set by the deviation of the area $A$ of a cell from a target value $A_0$, providing a 2D toy model of 3D cell volume incompressibility. The second contribution is set by the deviation of the cell perimeter $P$ from a target value $P_0$. Summing over all cells in the packing, the energy is then
\begin{equation}
\label{eqn:energy}
    E=\frac{1}{2}\sum_{n=1}^N \left[\kappa_A (A_n-A_{n0})^2 + \kappa_P (P_n-P_{n0})^2\right].
\end{equation}
 The quantity $p_0 = P_{n0}/ \sqrt{A_{n0}}$ defines a cell shape factor and is an important control parameter in our study. We set it the same for all cells in any simulation, independent of $n$.   In physical terms, $p_0$ is commonly attributed to a competition between cell cortical contractility and cell-cell adhesion ~\cite{Staple2010,Nagai2001,Farhadifar2007,Bi2016,Bi2015}, although recent experiments also imply a relationship with cell-substrate traction \cite{saraswathibhatla2020tractions}. Cell shape has been shown experimentally to predict jamming behaviour in epithelial tissues \cite{Park_NMAT_2015,Grosser2021}. The elastic constants $\kappa_P$ and $\kappa_A$ set the strength of the perimeter and area interactions, and we choose $\kappa_P=1$ as our basic unit of stress.

 The elastic forces exerted on the vertices of any cell due to the elastic contributions of that cell are sketched in Fig.~\ref{fig:force_diag}(a).  Consider in this sketch a representative edge of length $L$, connecting two representative adjacent vertices. The cell that is sketched then contributes to each of these two vertices an equal and opposite tension-like force of  magnitude $\kappa_{\rm P}(P-P_0)$, acting tangentially along the edge, inwards along the edge when $P > P_0$, and outwards when $P < P_0$.  The cell shown also contributes to the same two vertices a pressure-like force  of magnitude $\kappa_{\rm A}(A-A_0)L$,  acting perpendicularly to the edge, in towards the cell when $A > A_0$, and outwards when $A < A_0$.   These expressions are derived in the Appendix.
Each vertex in Fig.~\ref{fig:force_diag}a) additionally belongs to two further cells (not shown) that contribute corresponding elastic forces. The total elastic force $\Vec{F_j}$ on the $j$th vertex in the tiling is calculated by summing these contributions from its three shared cells.  

In an externally applied simple shear flow of rate $\gdot$, with  flow direction $x$ and flow-gradient direction $y$, the position $\Vec{r_j}$ of the $j$th vertex in the tiling obeys over-damped dynamics with drag coefficient $\zeta$:
\begin{equation}
    \frac{d\Vec{r_j}}{dt} =\frac{1}{\zeta} (\Vec{F_j}+ v\sum_{i_j = 1}^{3}{w_{i_j}\Vec{\hat{n}}_{i_j}}) + \gdot y_j \Vec{\hat{x}},
    \label{eq:motion}
\end{equation}
with Lees-Edwards periodic boundary conditions \cite{allen2017computer}.

\begin{table}[t]
\caption{Parameters of the model. Dimensions are expressed in terms of modulus (G), time (T) and length (L).}
\label{tb:independants}
\begin{tabular}{llll}
\hline
Quantity                                                               & Symbol    & Dimensions & Value            \\ \hline
Number of Cells                                                        & N         & 1          & 100              \\
Drag coefficient                                                       & $\zeta$      & GT        & 1 (time unit)    \\
\begin{tabular}[c]{@{}l@{}}Edge length \\ at initialisation\end{tabular}  & L         & L          & 1 (length unit)  \\
Perimeter modulus                                                      & $\kappa_{\rm P}$   & G          & 1 (stress unit) \\
Area modulus                                                           & $\kappa_{\rm A}$   & GL$^2$       & 1/2                \\
Bidispersity                                                           & -         & 1          & 1 : 1.4          \\
\begin{tabular}[c]{@{}l@{}}Polarisation angle \\ diffusion\end{tabular} & $D_r$      & T$^{-1}$     & 0.5              \\
T1 threshold                                                        & $l_{T1}$         & L          & 0.07              \\
Shape factor                                                           & $p_0$      & 1          & Varied           \\
Activity                                                               & $v$         & GL         & Varied           \\
Shear rate                                                             & $\dot{\gamma}$ & T$^{-1}$   & Varied           \\ 
Time step                                                         & 
$dt$    & T       & 0.01         \\
\hline
\end{tabular}
\end{table}

The second term on the right hand side of this equation describes a random motile activity{~\cite{Bi_PRX_2016,Das_neighbor_exchange,D3SM00327B}. The magnitude $v$ of this activity is an important control parameter in our study.  The direction of the motility of the $j$th vertex in the tiling is prescribed by the weighted sum of the polarisation vectors $\Vec{\hat{n}}_{i_j}=(\cos \theta_{i_j},\sin \theta_{i_j})$ of the three cells $i_j=1,2,3$ in contact with that $j$th vertex. The polarisation angle of each cell in the tiling is initialised at the start of any simulation randomly from a top-hat distribution in the range $0$ to $2\pi$. It thereafter experiences angular diffusion with a diffusion coefficient $D_r$, modelled via Gaussian random noise.  Accordingly, 
the polarisation angle of the $n$-th cell in the tiling obeys
\begin{equation}
\frac{d\theta_n}{dt} = \eta_n,
\end{equation}
in which $\eta_n$ is a random variable with statistics~\cite{Bi2016}
\begin{equation}
    \langle \eta_n(t)\rangle = 0,\;\;\; \langle\eta_n (t)\eta_m (t')\rangle = 2D_r\delta_{nm}\delta(t-t').
\end{equation}

The weighting factors $w_{i_j}$ in Eqn.~\ref{eq:motion} ensure that the largest contribution to the polarisation vector of our representative vertex (the $j$th in the tiling) arises  from whichever of its three associated cells $i_j=1,2,3$ has the largest value of the summed lengths of cell edges that contact that vertex. Specifically, we define $l_{i_j}$ to be the summed length of the two edges of the $i_j$th cell in contact with vertex $j$, as shown by the colour coded lines in Fig.~\ref{fig:force_diag}b), and set
\begin{equation}
w_{i_j} = \frac{l_{i_j}}{12L_{Tj}},
  \label{eq:weighting}
\end{equation}
consistent with the weighting function used in previous work ~\cite{Das_neighbor_exchange,Mitchel2020}. Here $L_{Tj}$ is the total length of the three edges in contact with vertex $j$. 
Topological T1 cell-cell rearrangement events also intermittently arise, leading to plastic stress relaxation. Specifically, when any cell edge length becomes smaller than a threshold value $l_{T1}$, a T1 event occurs.  Prior to a T1, the selected edge is defined by two vertices, one shared between cells $\alpha \beta \gamma$ and the other $\alpha \beta \delta$. The T1 event then replaces these two old vertices with  two new ones, shared by cells $\alpha \gamma \delta$ and $\beta \gamma \delta$ \cite{Bi2014,Mitchel2020}. 

To initialise an amorphous cellular tiling, we start from a uniform lattice of monodisperse hexagonal cells of cell edge length $1$ stacked in $\sqrt{N}$ rows, each of $\sqrt{N}$ cells. Target perimeter and area values are then assigned to each cell. To avoid the effects of crystallisation associated with monodisperse packings \cite{Mari2014} we use a bidisperse packing  in which half the cells have a smaller size and half a larger size. Specifically, we assign these  two populations target perimeters in ratio $ 1:1.4 $  respectively. To maintain a consistent target shape factor ($p_0=\frac{P_0}{\sqrt{A_0}}$) between these two populations, their target areas are set in ratio $1:1.4^2$ respectively}.  The overall scale of target area is set such that the target area summed over all cells equals that of the domain size created in the initial uniform hexagonal tiling. The packing is then randomised by implementing cell motility with non-zero $v=v_{\rm prep}$ and $D_r=D_{r, \rm prep}$ in the absence of shear for $t_{\rm prep}$ time units then subsequently relaxing the system with zero activity (and zero shear) for $t_{\rm relax}$ time units. Choosing $v_{\rm prep}=4.0,  D_{r, \rm prep}=0.25,t_{\rm prep}=14.5,t_{\rm relax}=5.5$ was found to produce a random bidisperse initial cellular tiling with fully relaxed cell areas and perimeters. 

The equations of motion described above are integrated using the explicit Euler method with time step $dt$, both during the preparation stage just discussed and the subsequent shearing stage. The timestep $dt$ used was converged to the limit $dt\to 0$, and the system size was converged to the limit $N\to\infty$.

The values, symbols and dimensions for the parameters of the model and shear protocol are listed in Table~\ref{tb:independants}.

In what follows, we report the steady state shear stress $\sigma$ in the tissue as follows. For any individual cell with vertices numbered $c=1\cdots C$, a cell level stress is calculated at any time as~\cite{Andia2006}
\begin{equation}
S_{\alpha \beta} = \frac{1}{A}\sum_{c=1}^{C}{F}_{c\alpha} {r}_{c\beta}.
\label{eq:stress}
\end{equation}
Here $F_{c\alpha}$ is the $\alpha$ component of the elastic force on vertex $c$ from the elastic contributions of that single cell,  and $r_{c\beta}$ is the $\beta$ component of position of vertex $c$. $A$ is the cell's area. These individual cell stresses, thus defined, are then averaged over all cells in the packing. The packing-averaged shear stress is then averaged over many strain units once a state of statistically steady shear is attained, and furthermore (for each set of model parameters) over three runs with different random number seeds. It is this averaged shear stress that is reported in the results that follow. We have checked it to be robust to changes in system size for $N>100$. Fluctuations about the average decrease with increasing $N$.

In the absence of internal activity ($v=0$) and external applied shear ($\gdot=0$), the vertex model captures a fluid-solid transition at a critical target cell shape $p_0 = p_0^*$
~\cite{Bi2015,Bi2016,lin2019dynamic,Yan2019}, with $p_0^*\approx 3.81$ for the bidisperse tiling studied here.  For $p_0 < p_0^*$, cells cannot attain their target shape and the energy barriers to local T1 cell rearrangements are significant: the tissue resists shear, giving a solid phase. For $p_0 > p_0^*$, cells do achieve their target shape and the energy barriers to rearrangements are small, resulting in a liquid phase that cannot resist shear~\cite{Bi2015}. A nonlinear shear applied quasistatically ($\gdot\to 0$) however induces a solidification transition at a critical strain $\gamma_{\rm c}(p_0)$ for  $p_0^*<p_0<p_0^{**}$, with $p_0^{**}\approx 4.03$~\cite{Cochran2021}. It does so by deforming cells such that they can no longer attain their target shape, eliminating the zero-shear liquid and inducing a solid-like response. The steady state flow curve of shear stress vs shear rate, $\sigma(\gdot)$, then displays a yield stress $\sigma_{\rm Y}=\lim_{\gdot\to 0}\sigma(\gdot)\neq 0$  for all $p_0<p_0^{**}$~\cite{Cochran2021}.

\section{Results}
\begin{figure}[!t]
\centering
\includegraphics[width=0.23\textwidth,height=0.32\textwidth]{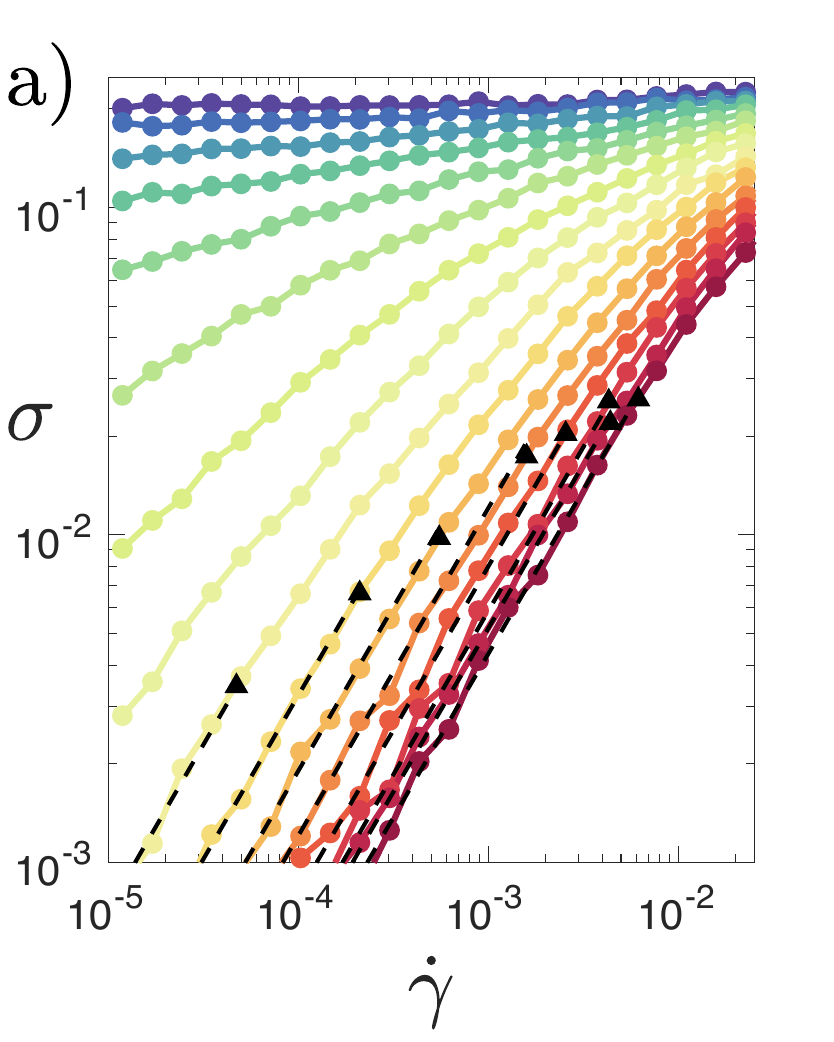}    \includegraphics[width=0.23\textwidth,height=0.32\textwidth]{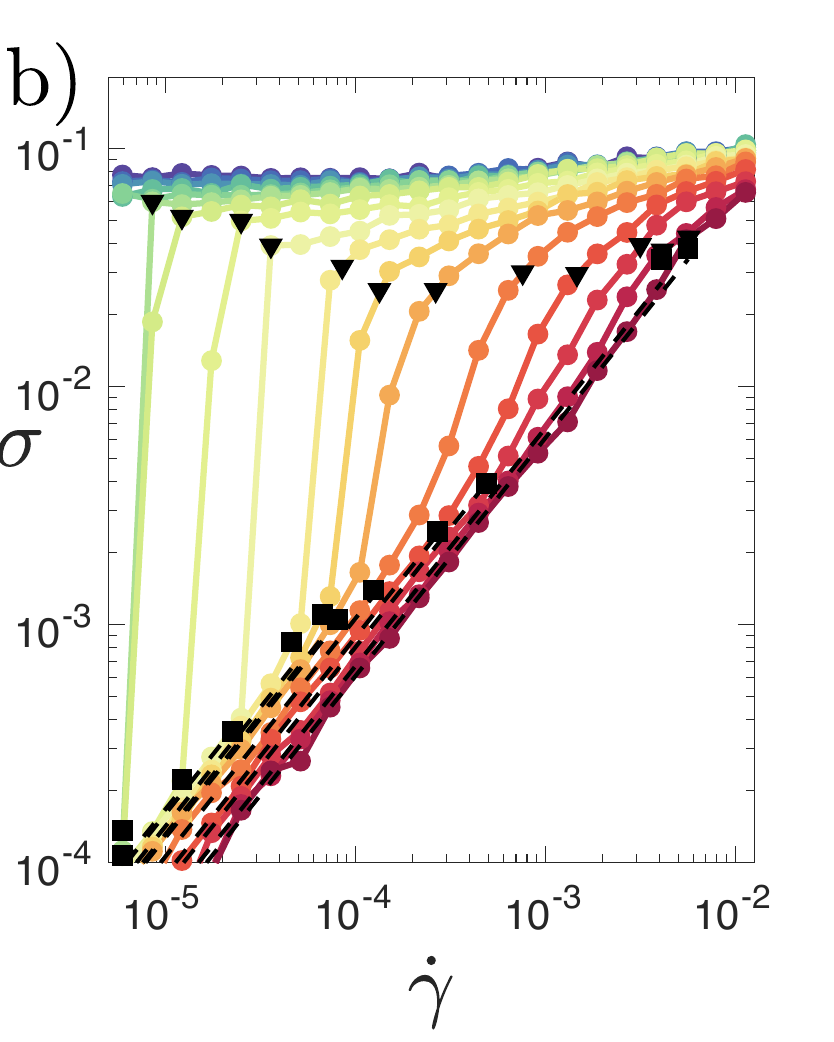}
\caption{ Steady state flow curves of shear stress as a function of shear rate. {\bf a)} For fixed $p_0 = 3.65$ in the (zero activity, zero shear) solid phase with activity $v$ = 0.0, 0.1, 0.2, 0.3, 0.4, 0.5, 0.6, 0.7, 0.8, 0.9, 1.0, 1.1, 1.2, 1.3, 1.4, 1.5 in curves downwards blue to red. For low activity, we find a yield stress in the limit of low strain rate. For high activity, we see Newtonian flow response at low strain rates with shear thinning for higher strain rates. {\bf b)}  For a fixed $p_0 = 3.90$ in the (zero activity, zero shear) liquid phase with activity $v$ = 0.00, 0.01, 0.02, 0.03, 0.04, 0.05, 0.06, 0.08, 0.10, 0.12, 0.14, 0.16, 0.20, 0.25, 0.30, 0.35, 0.40 in curves downwards blue to red. With no activity, we find a yield stress in the limit of low strain rate. With modest levels of activity, Newtonian flow response at low strain rate gives way to a discontinuous shear thickening transition with increasing shear rate.  Dashed lines fit regimes of constant viscosity,  $\eta(p_0,v)=\sigma/\gdot$. Black squares show  $\gdot_{\rm thick}$ and triangles $\gdot_{\rm thin}$, defined in the main text.}
\label{fig:Flow_curves}
\end{figure}

We start by exploring the effects of activity on a sheared tissue, reporting in Fig.~\ref{fig:Flow_curves}a) steady state flow curves $\sigma(\gdot)$ in the (zero-activity, zero-shear) solid phase, $p_0<p_0^*$. At zero activity we see a yield stress, $\sigma_{\rm Y}=\lim_{\gdot\to 0}\sigma(\gdot)\neq 0$, indicating a solid-like response with infinite viscosity $\eta=\sigma/\gdot$ in quasistatic shear $\gdot\to 0$, consistent with~\cite{Cochran2021}. In contrast, at high activity we find liquid-like flow with $\sigma=\eta \gdot$, in the limit of small shear rate  $\gdot\to 0$. This is termed Newtonian flow behaviour. The viscosity  $\eta=\eta(p_0,v)$ is fit by the black dashed lines.
 
 \begin{figure}[t]
{
    \includegraphics[width=0.40\textwidth]{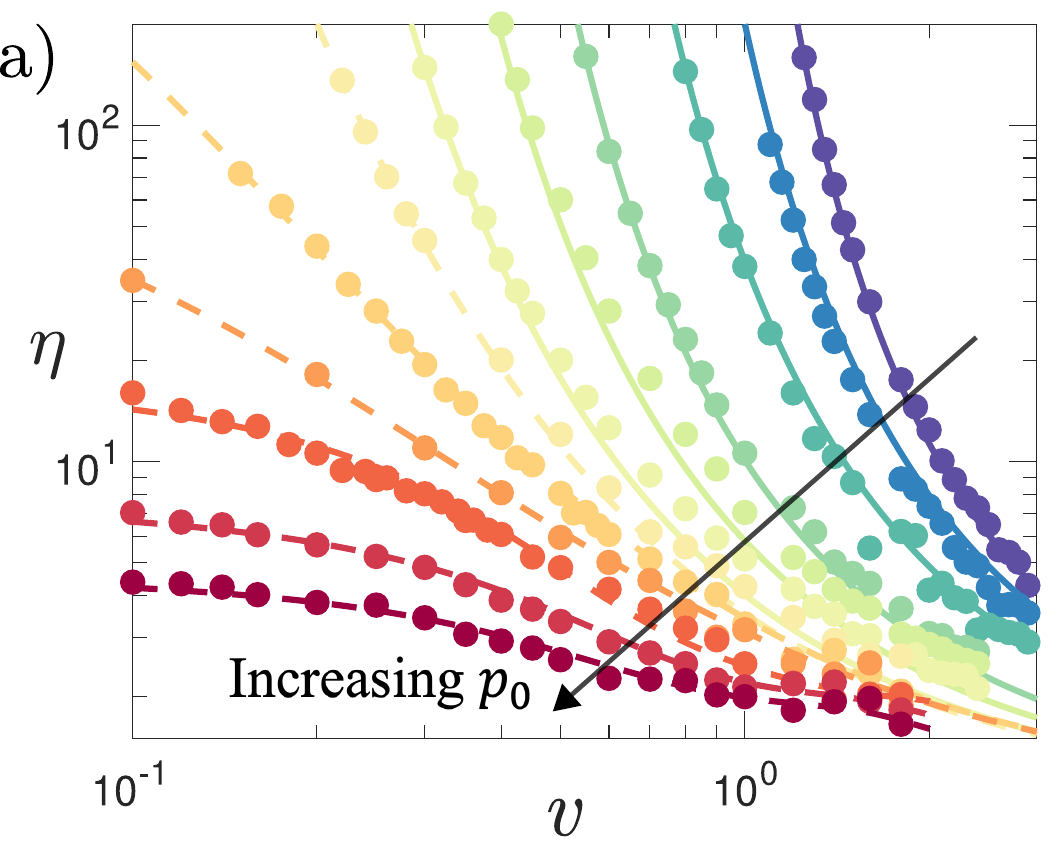}
    \includegraphics[width=0.425\textwidth]{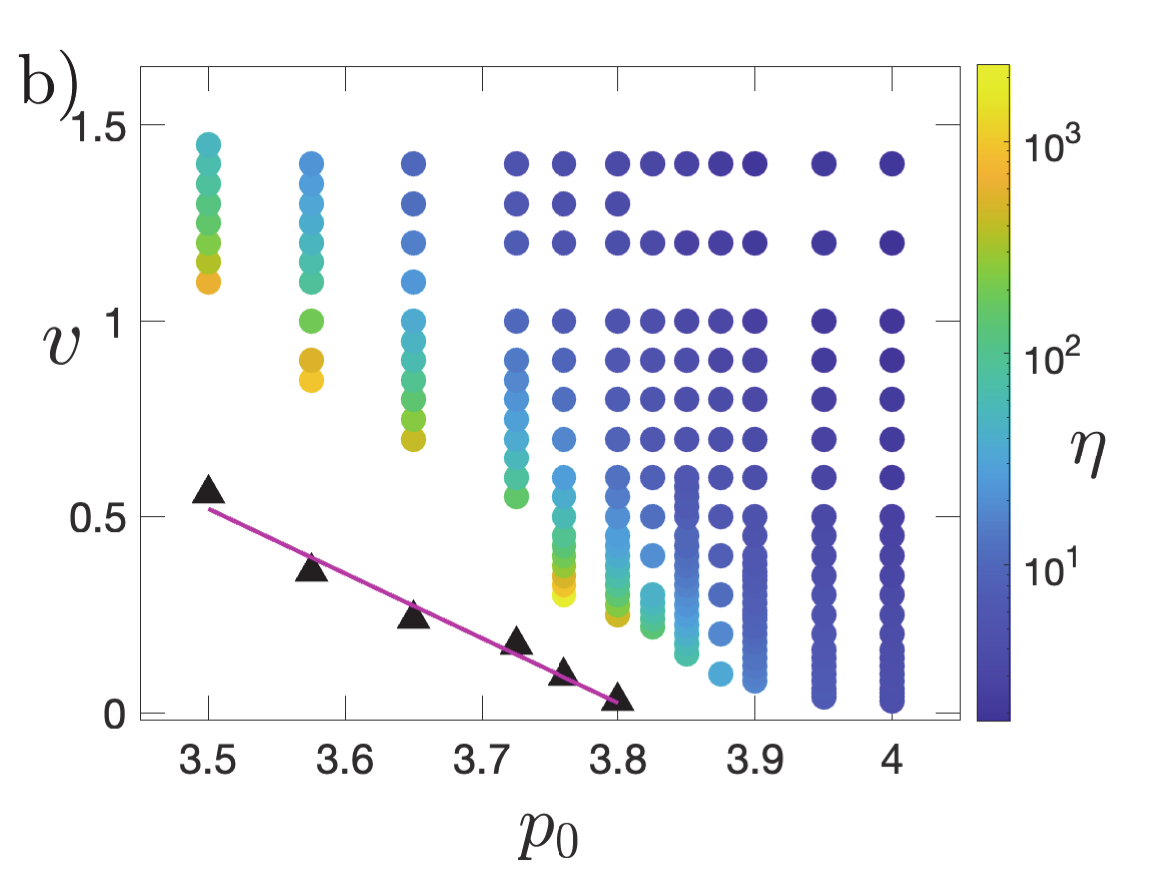}
}
\caption{Newtonian viscosity $\eta(p_0,v)=\sigma/\gdot$ from black-dashed fits in Fig.~\ref{fig:Flow_curves},  
{\bf a)} Plotted vs. activity  $v$ for target shape $p_0 = $ 3.500, 3.575, 3.650, 3.725, 3.760, 3.800, 3.825, 3.850, 3.875, 3.900, 3.950, 4.000 in curves downwards blue to red. Solid lines: fits to Vogel-Fulcher-Tammann (VFT) form $\eta\sim \exp\left[1/(v-v_{\rm c})\right]$ for $p_0<p_0^*\approx3.81$, suggesting a viscosity divergence as $v\to v_{\rm c} (p_0)$. Dashed lines: spline fits for $p_0>p_0^*$. 
{\bf b)} Plotted as a colourmap in the plane of $v$ and $p_0$. Black triangles show the value of $v=v_{\rm c}(p_0)$ at which the VFT fit predicts the viscosity to diverge. Magenta line: linear fit to black triangles.
}
\label{fig:viscosity2}
\end{figure}
\begin{figure}[!t]
\centering
    \includegraphics[width=0.235\textwidth]{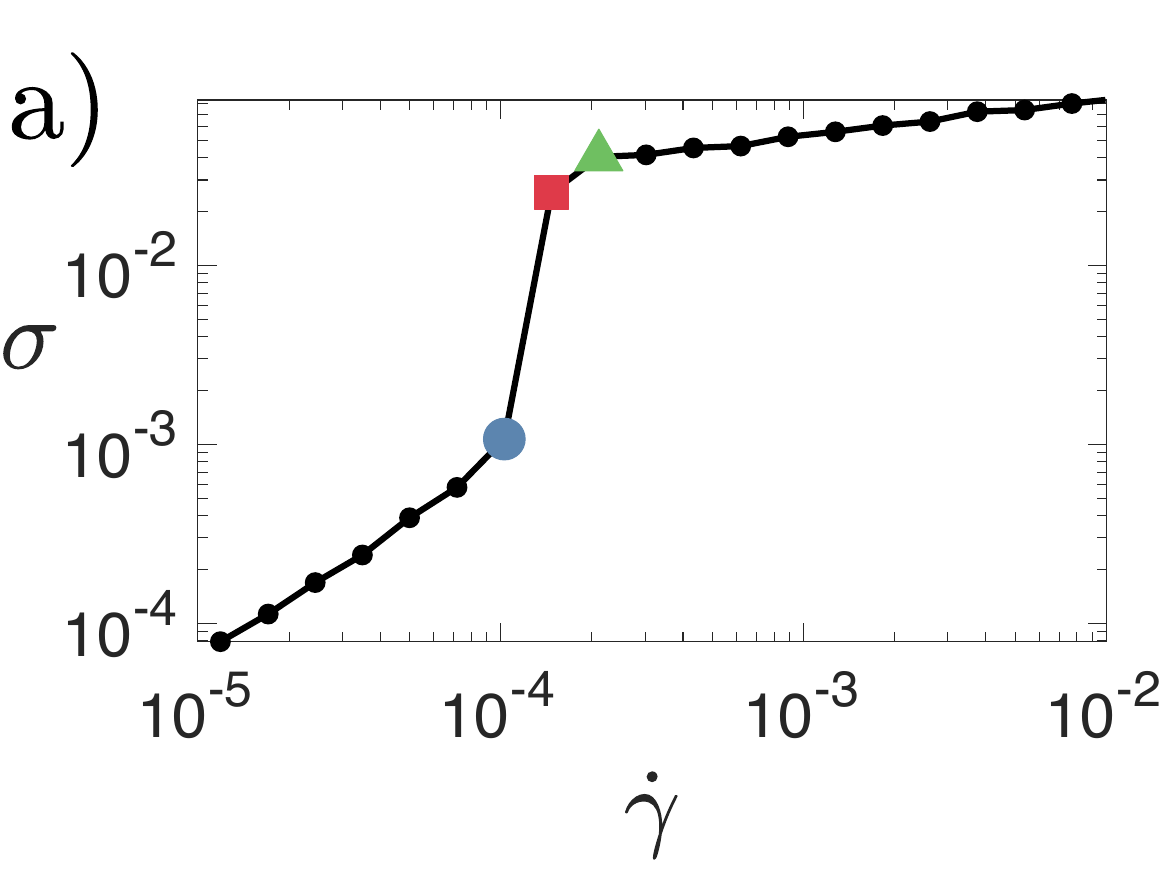}
    \includegraphics[width=0.235\textwidth]{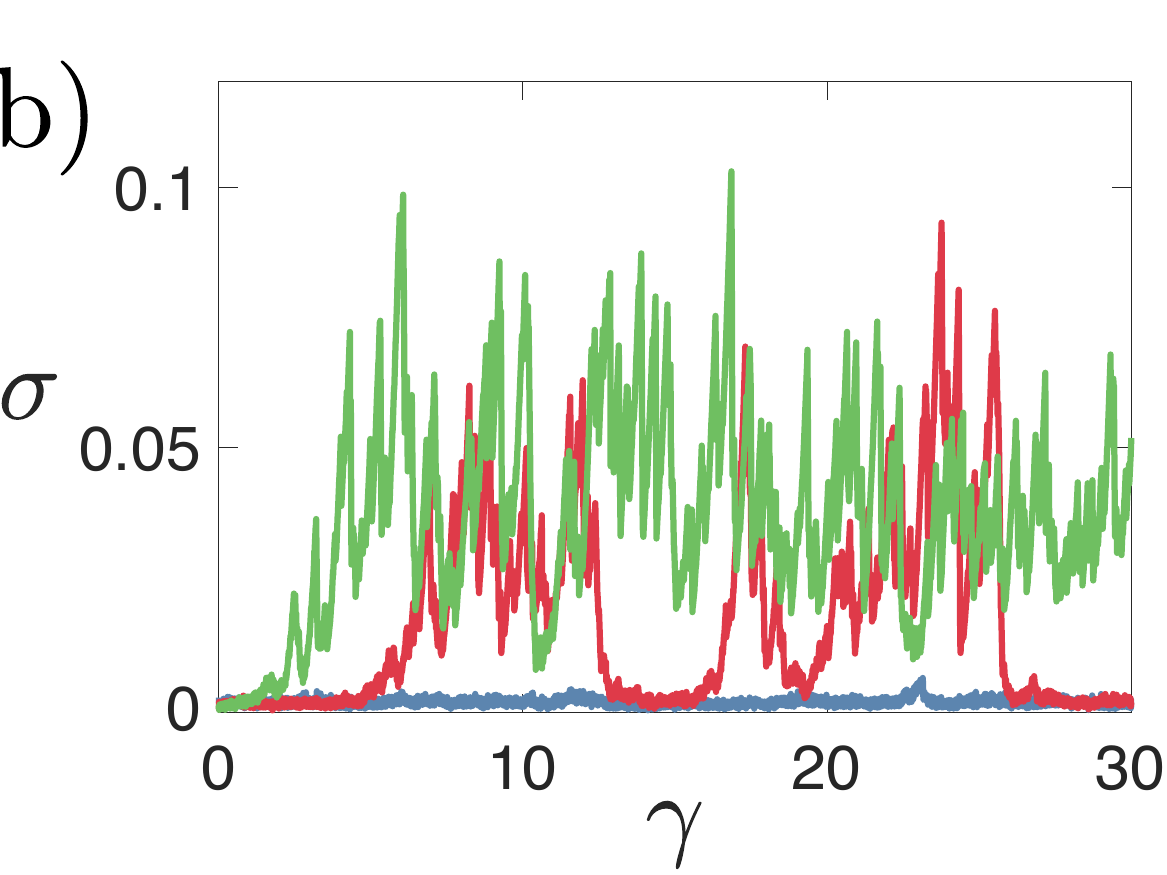}
    \includegraphics[width=0.235\textwidth]{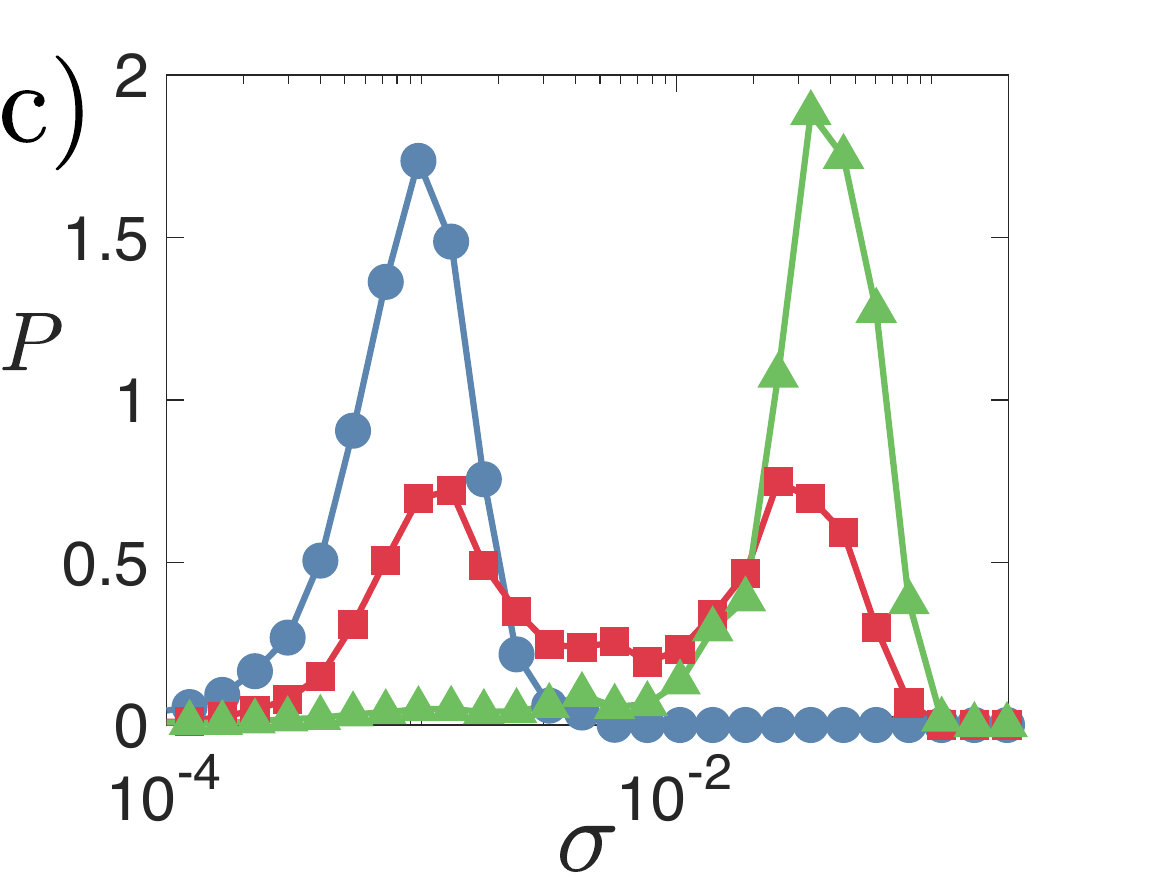}
    \includegraphics[width=0.235\textwidth]{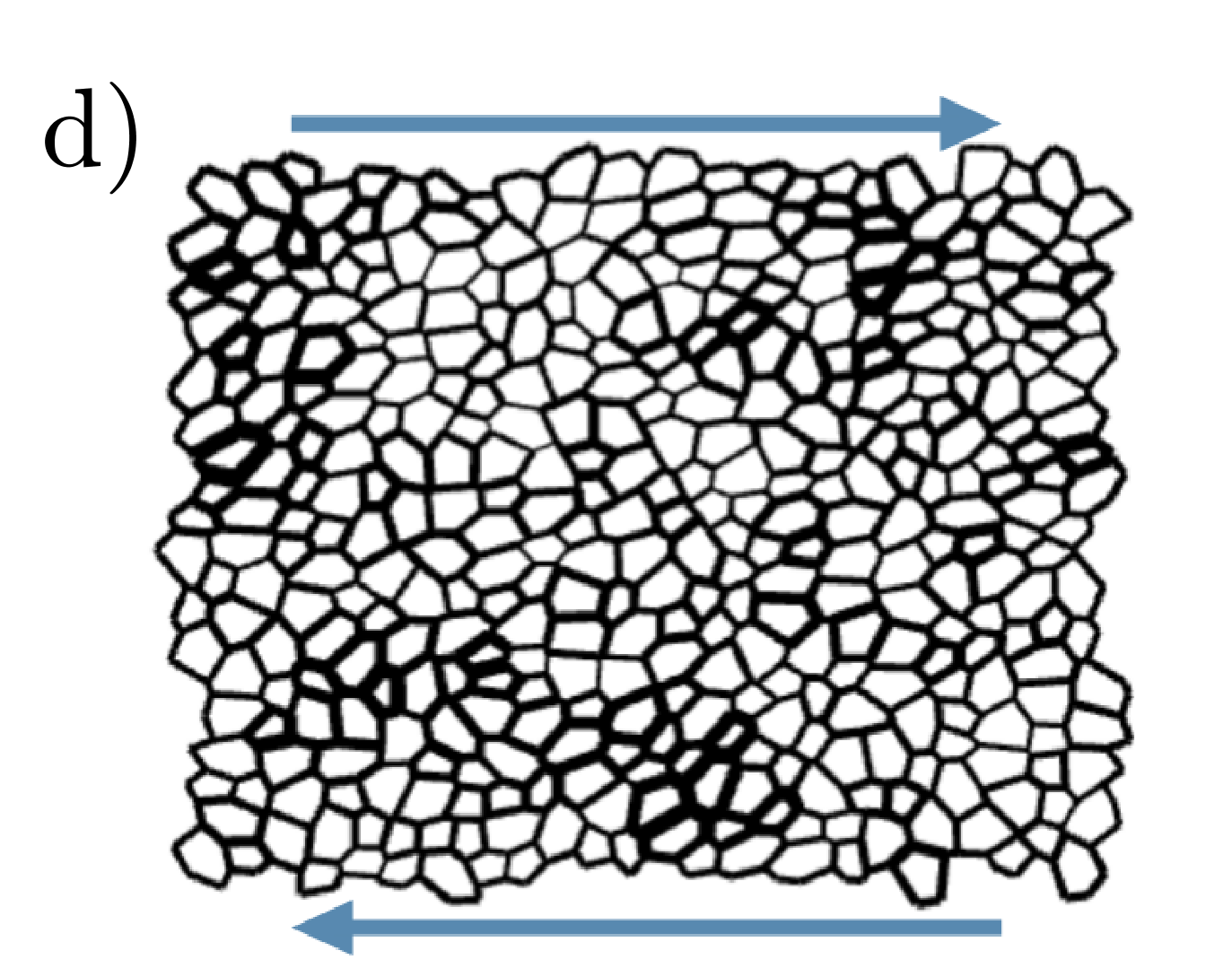}
    \includegraphics[width=0.235\textwidth]{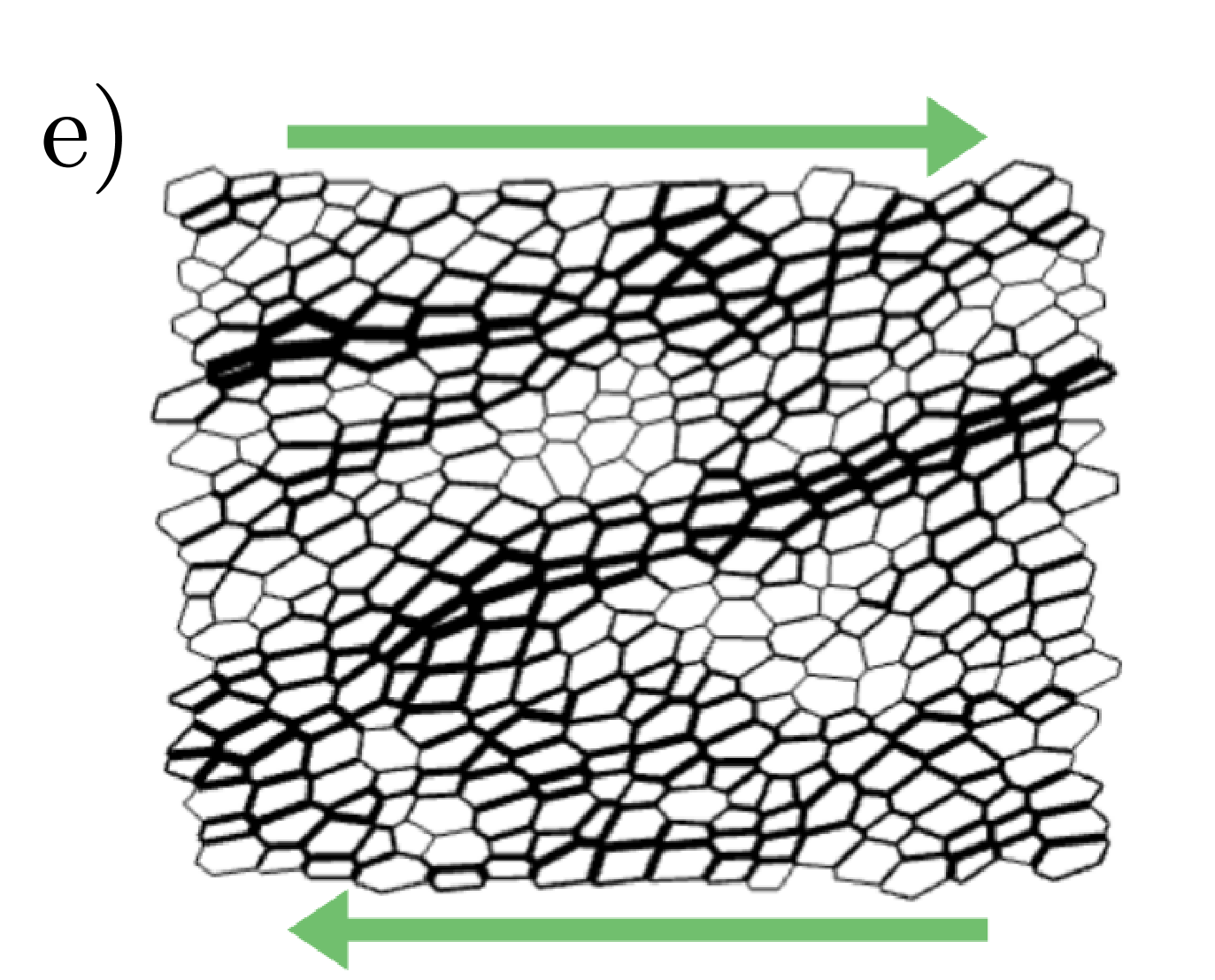}
 \caption{Exploring the discontinuous shear thickening (DST) transition. {\bf a)} Representative flow curve showing the DST transition. {\bf b)} Stress as a function of strain $\gamma=\gdot t$ for the three imposed strain rates denoted by shapes of corresponding colour in a): Blue circles $\dot{\gamma}$ = 5.14x10$^{-5}$, red squares $\dot{\gamma}$ = 7.36x10$^{-5}$ and green triangles $\dot{\gamma}$ = 1.1x10$^{-4}$.  {\bf c)} Corresponding probability distributions of the logarithm of the stress. Representative state snapshots at {\bf d)} $\dot{\gamma}=5.14$x$10^{-5}$ (blue circles) and {\bf e)} $\dot{\gamma}=1.1$x10$^{-4}$ (green triangles), with the line thickness of any cell edge proportional to the  tensile stress it carries. Regions of high stress are distributed through the system in d), but formed into system-spanning force chains in e). Target call shape $p_0=3.9$, activity $v=0.12$.
 }
 \label{fig:DST}
\end{figure}

The zero-shear ($\gdot\to 0$) viscosity $\eta(p_0,v)$ in the (zero-activity, zero-shear) solid phase, $p_0<p_0^*$,  thus increases dramatically with decreasing activity $v$, at fixed target cell shape. In Fig.~\ref{fig:viscosity2}a) we fit this increase to the Vogel-Fulcher-Tamman (VFT) form $\eta\sim \exp\left[1/(v-v_{\rm c})\right]$ 
to find the critical activity $v=v_{\rm c}(p_0)>0$ below which $\eta$ diverges, at any $p_0<p_0^*$. This divergence of the zero shear viscosity at a critical $v_{\rm c}(p_0)$ for $p_0<p_0^*$ may indicate a true yield stress $\sigma_{\rm Y}$ for all $0\le v<v_{\rm c}$, consistent with that at  $v=0$~\cite{Cochran2021}, although a power law $\sigma\propto\gdot^n$ with $n<1$ is not ruled out for $0< v<v_{\rm c}$. Either way, in the  solid phase, $p_0<p_0^*$, a critical activity $v_{\rm c}(p_0)$ is needed to eliminate solid-like behaviour  in favour of Newtonian flow with finite $\eta$. This critical $v_{\rm c}$ is plotted vs. $p_0$ in Fig.~\ref{fig:viscosity2}b), which also shows a colour map of $\eta$ in the plane of $v,p_0$. The zero-shear viscosity is also consistent with the viscosity calculated based on the Green-Kubo relation~\cite{Yang2017}. A linear fit suggests that $v_{\rm c}$ falls to zero at the (zero activity, zero shear) solid-liquid transition $p_0=p_0^*$. This intercept is consistent with similar data in linear studies \cite{Yang2017}. However, the curve shape differs, implying a different mechanism at non-zero activity. 
 
 The flow curves  just discussed, for a tissue in its (zero-activity, zero-shear) solid phase, closely resemble those of complex fluids such as glassy colloidal and jammed athermal soft particle suspensions~\cite{Bonn2017,Petekidis2004,Hunter2012}. These show a yield stress at high packing fraction $\phi$ and low temperature, analogous to our curves for low activity. Particle suspensions also show low shear Newtonian behaviour at low $\phi$ and high temperature, analogous to ours at high activity. 
 
We next consider the effect of activity on a sheared tissue in its liquid phase, $p_0>p_0^*$ See the steady state flow curves in Fig.~\ref{fig:Flow_curves}b. In notable contrast to the solid phase, these closely resemble the flow curves of dense frictional suspensions and granular matter~\cite{Wyart2014,Prabhu2022,Browne2014}. In particular,  they show {\em discontinuous shear thickening} (DST), in which the shear stress jumps discontinuously with increasing strain rate at high $\phi$ (in suspensions) or low activity (here). DST then gives way at lower $\phi$ (in suspensions) or higher activity (here) to continuous shear thickening (CST), in which the stress still steepens with shear rate, but without jumping.

\begin{figure}[!t]
\includegraphics[width=0.45\textwidth]{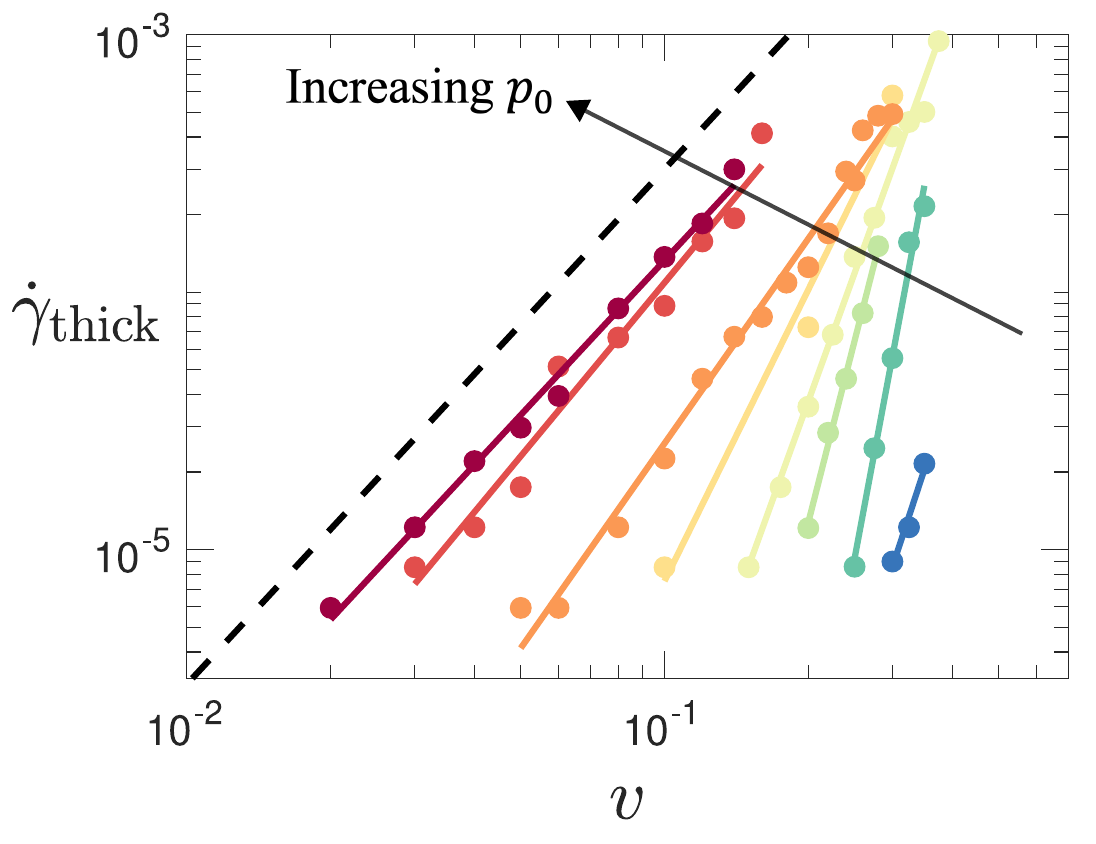}
\caption{Shear rate $\gdot_{\rm thick}$  at the onset of shear thickening as a function of activity for target cell shape $p_0=$   $3.760, 3.800, 3.825, 3.850, 3.875, 3.900, 3.950, 4.00$ in curves blue to red leftwards. Coloured straight lines show power law fits to data, $\gdot_{\rm thick}\propto v^\alpha$. This implies that shear thickening will be present even at very low levels of activity. The dotted black line shows the power $\alpha = 2$ as a guide to the eye}.
\label{fig:gdotThicken}
\end{figure}

We next explore the origins of DST by examining the spatial-temporal evolution of the stress for different shear rates at fixed activity and $p_0$. An example flow curve for $v=0.12$, $p_0 = 3.90$ is shown in Fig.~\ref{fig:DST}a).  In Fig.~\ref{fig:DST}b), the  stress as a function of strain  $\gamma=\gdot t$ (which is proportional to time $t$, given constant $\gdot$) is plotted for three different shear rates, with line colours corresponding to marker colours in Fig.~\ref{fig:DST}a). The corresponding stress distributions over each of these strain series is shown in Fig.~\ref{fig:DST}b).  For low shear rate (blue),  the stress fluctuates modestly (in proportional terms) around a low value. Similarly for high shear rate (green) the stress fluctuates modestly around a high value. In contrast, at intermediate shear rate (red), the stress intermittently switches between these low and high stress states to give a bimodal distribution.

This intermittent, bimodal stress evolution at the DST transition is also seen in frictional suspensions, where it is caused by percolating compressive force chains~\cite{Sedes2020,Wyart2014,Cates1998,Seto2013,Behringer2019,Bi2011,Singh2020}. Our simulations likewise evidence percolating force chains in this vertex model of biological tissue: Fig.~\ref{fig:DST}c) and d) show representative state snapshots corresponding to the lowest and highest strain rates in Fig.~\ref{fig:DST}a,b), with the thickness of each cell edge proportional to the tensile stress it carries. In the low stress (unthickened) state, the tension is distributed fairly evenly across the system. In contrast, the high stress (thickened) state displays system-spanning force chains. In important contrast to the \emph{compressive} force chains that form in frictional suspensions, however, we  find these stresses to be tensile in nature in tissues. This is consistent with recent computational  ~\cite{noll2017active,Li_tension,Atia2018GeometricJamming} and experimental~\cite{Park_NMAT_2015, tambe2011collective} studies, which indeed found tensile stresses overwhelmingly to dominate in biological tissues.

To further characterise the regime of shear thickening, we define two shear rates, each via the logarithmic slope of the flow curve, $G=d\log\sigma/d\log\gdot$. ($G=1$ indicates Newtonian behaviour.) First, we define the onset of shear thickening in Fig.~\ref{fig:Flow_curves}b) via the shear rate $\gdot_{\rm thick}$ (shown by black squares) at which  $G$ first increases above $1+\epsilon$, with $\epsilon=0.2$.  Second, we define the  reversion to shear thinning at higher strain rates via the shear rate $\gdot_{\rm thin}$ (black triangles) at which $G$ first falls below $1-\delta$, with $\delta=0.1$.  At low activity, where DST arises, $\gdot_{\rm thin}=\gdot_{\rm thick}$, to within the resolution of $\gdot$ values simulated. 

\begin{figure}[!t]
\includegraphics[width=0.48\textwidth]{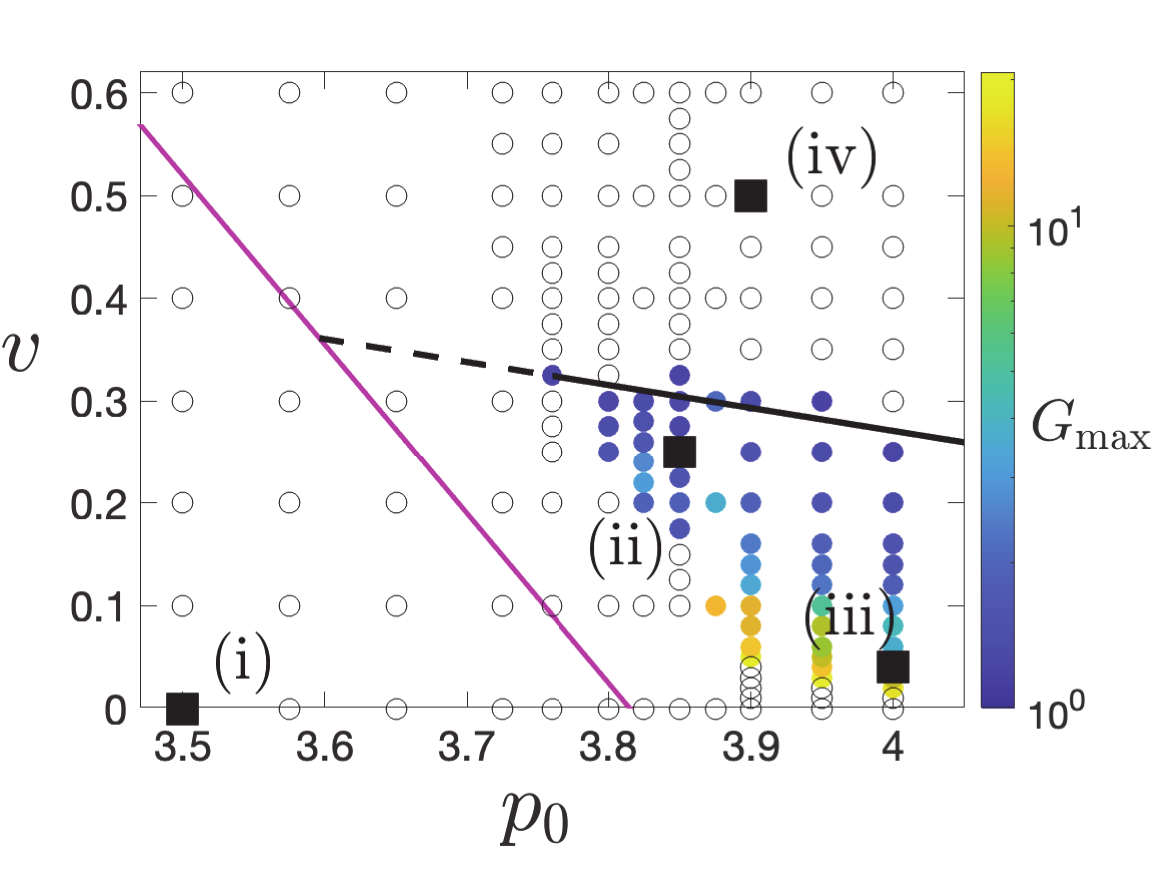}
\includegraphics[width=0.2\textwidth]{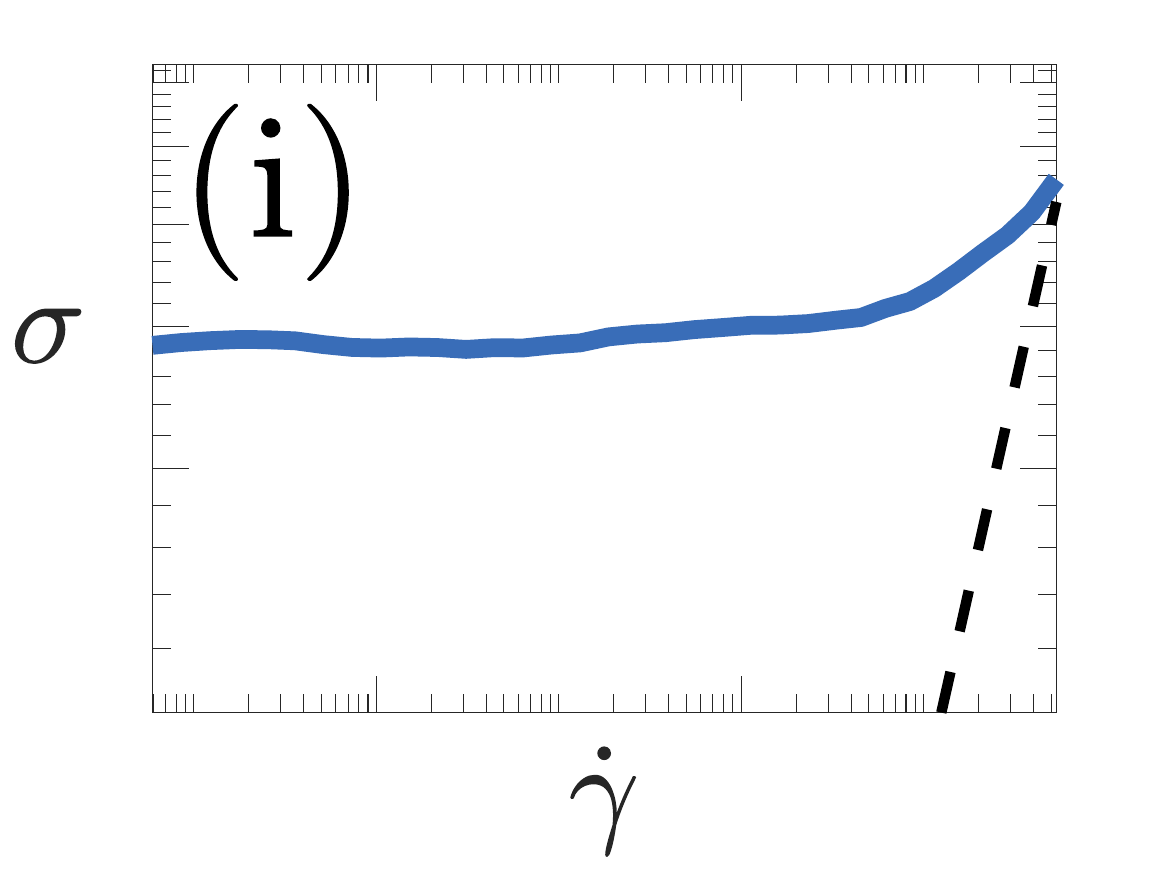}
\includegraphics[width=0.2\textwidth]{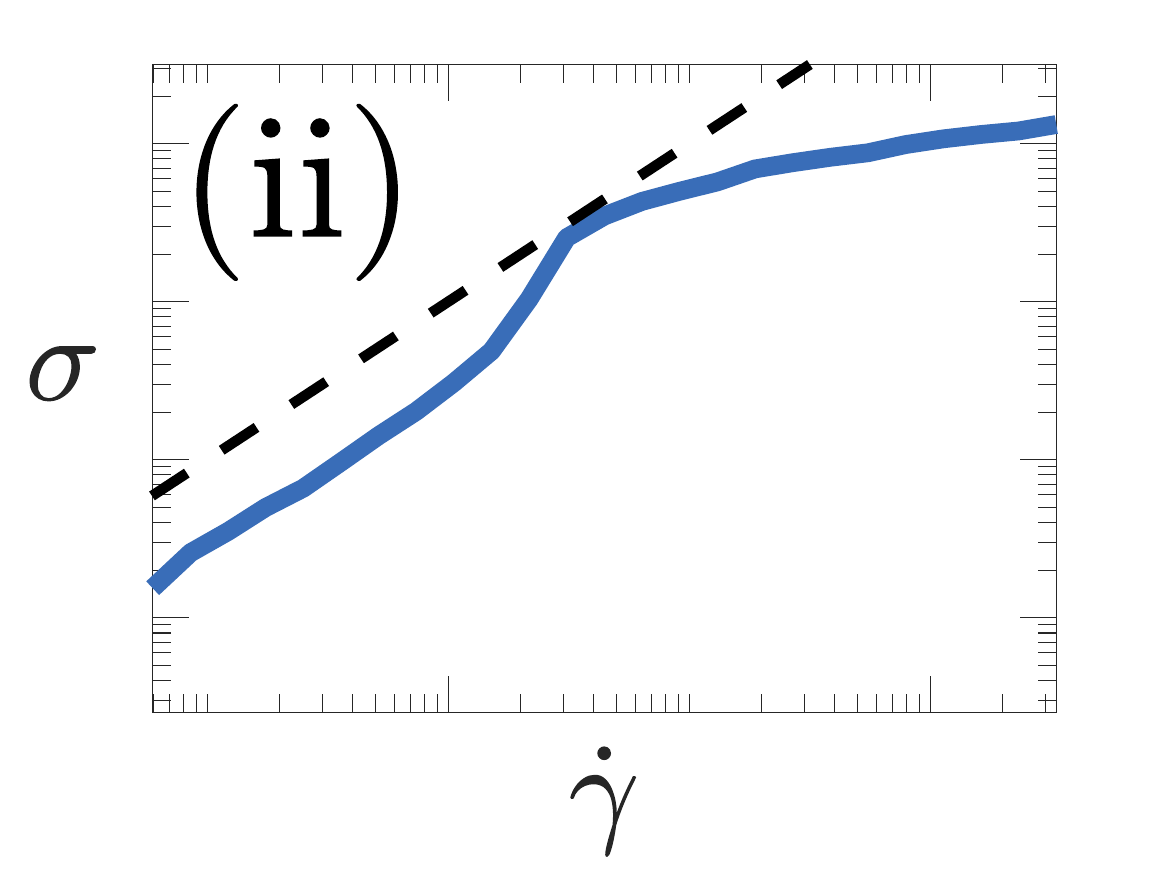}\\
\includegraphics[width=0.2\textwidth]{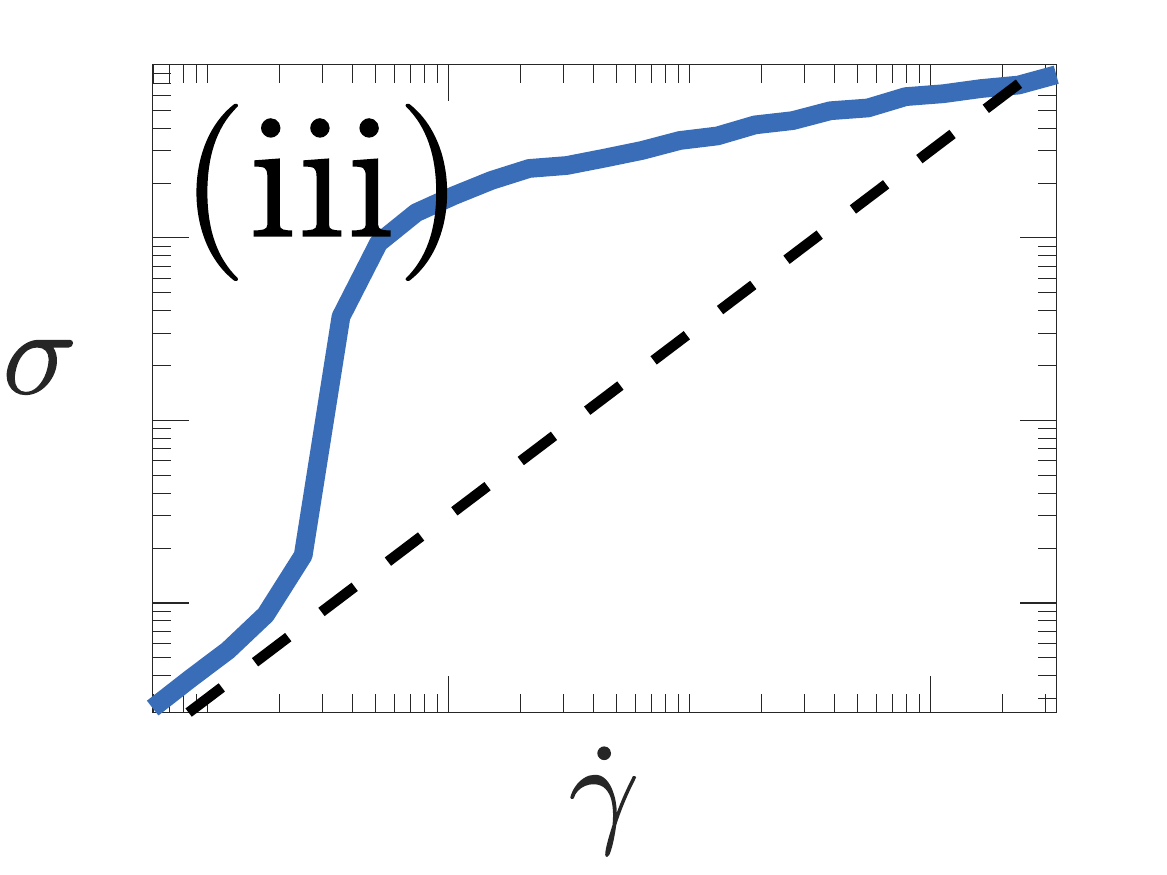}
\includegraphics[width=0.2\textwidth]{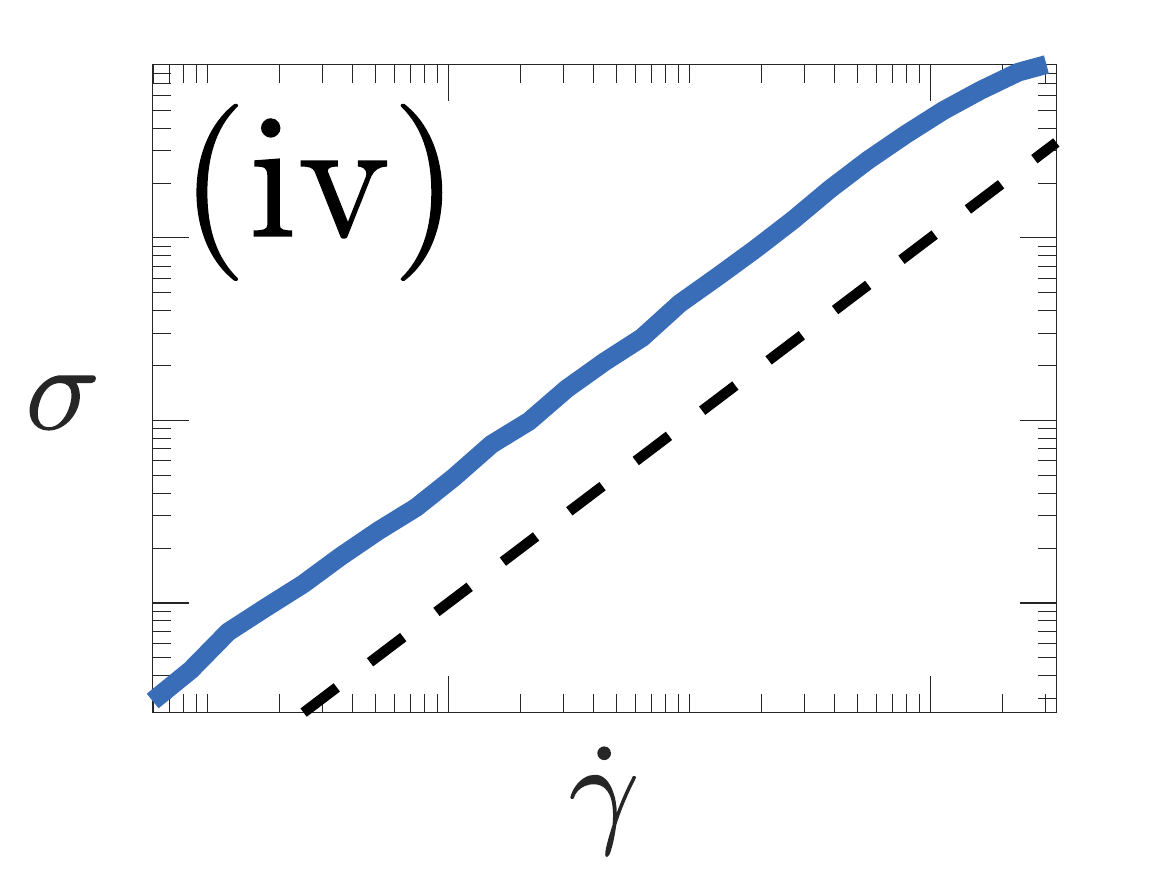}
\caption{Phase diagram showing different regimes of flow behaviour for different values of $v, p_0$, with a representative flow curve in each regime.
Top: Coloured symbols: maximum logarithmic slope $G_{\rm max}$ of the flow curve at any $v,p_0$, provided $G_{\rm max}>1+\epsilon$ with $\epsilon=0.2$, designated as the criterion for shear thickening. Open circles have $G_{\rm max}<1+\epsilon$, and no thickening. Black solid line: linear fit to $v=v(p_0)$ at which thickening is lost. Dashed line: extrapolation of black solid line left to meet magenta line. Magenta line as in Fig.~\ref{fig:viscosity2}. Panels {\bf i-iv)} show representative flow curves at the $v,p_0$ values indicated. {\bf i)} Yield stress, {\bf ii)} CST,  {\bf iii)} DST, {\bf iv)} Newtonian.}
\label{fig:phaseDiagram}
\end{figure}

Fig.~\ref{fig:gdotThicken}) shows $\gdot_{\rm thick}$ as a function of activity $v$ for several values of the target shape $p_0$. 
For $p_0$ values comfortably inside the (zero activity, zero shear) fluid phase above $p_0^*$, we find  $\gdot_{\rm thin}\approx\gdot_{\rm thick}\sim v^\alpha$ at low $v$. The exponent $\alpha$ decreases with increasing $p_0$, with $\alpha\approx 2.0$ at $p_0=4.0$. In the fluid phase $p_0>p_0^*$, therefore, {\em any} level of activity $v$, {\em however small}, is sufficient to restore Newtonian response $\sigma=\eta\gdot$ in quasistatic shear $\gdot\to 0$, as $\gdot < \gdot_{\rm thin}\approx\gdot_{\rm thick}\sim v^\alpha$. This contrasts notably with the (zero activity, zero shear) solid phase, $p_0<p_0^*$, where a finite level of activity $v=v_{\rm c}(p_0)$ is needed to give Newtonian behaviour in slow shear, $\gdot\to 0$. 

Having examined the shear rate at which shear thickening  (if present) arises in the flow curve for any pairing of $p_0,v$ values, we consider finally which flow curves indeed show shear thickening. To do so, we define $G_{\rm max}$ to be the maximum of the logarithmic gradient $G=d\log\sigma/d\log\gdot$ across each flow curve and plot in Fig.~\ref{fig:phaseDiagram} a colourmap of values of $G_{\rm max}$  that exceed $1+\epsilon$ with $\epsilon=0.2$, taking this as the minimal threshold for shear thickening. Values of $p_0,v$ for which the flow curve does not meet this threshold are shown as white open symbols. As can be seen, very strong shear thickening (large $G_{\rm max}$) arises at high $p_0$ and low $v$: this is the regime of DST, where the value of $G_{\rm max}$ is limited only by the resolution of $\gdot$ values simulated. As $v$ increases at fixed $p_0$,  we see a crossover to more moderate CST before thickening is lost at high $v$. The black solid line shows a linear fit to the threshold at which thickening is lost.

The apparent loss of shear thickening at fixed $v$ with decreasing $p_0$ in Fig.~\ref{fig:phaseDiagram} is worthy of comment. Towards the left hand edge of the regime of coloured symbols, the shear rate $\gdot_{\rm thick}$ that marks the onset of thickening falls to approach the minimum shear rate that we can feasibly simulate. Were we able to simulate arbitrarily low shear rates, we speculate that the observed regime of thickening would in fact extend leftwards, with $\gdot_{\rm thick}\to 0$ only at the magenta line, consistent with the zero-shear viscosity $\eta$ being infinite to the left of that line (Fig.~\ref{fig:viscosity2}).  
We have therefore continued the black solid line leftwards as a dashed line and suggest that the black and magenta lines together delineate the key rheological regimes observed in this work. Representative flow curves for each regime  are shown beneath the colourmap in Fig.~\ref{fig:phaseDiagram}. 

At higher shear rates, for all $p_0,v$, we observe shear thinning arising from T1 cell rearrangement events. This has been seen previously in a vertex model, and derives from an interplay of active fluctuations in vertex length with T1 transitions induced by shear~\cite{Duclut2021}.

\section{Discussions and Conclusions}
 Our work points towards a framework for understanding the emergent nonlinear mechanics  of biological tissue. In particular, we  have shown the nonlinear shear rheology of the vertex model  to be determined by an intricate interplay between the intrinsic solid-liquid transition that arises at a target cell shape $p_0=p_0^*\approx 3.81$ in the absence of shear or activity~\cite{Bi2016,Kranjnc2020}, with the mutually competing effects of a global external shear and local internal cell motility. 

Indeed, in slow shear, $\gdot\to 0$, a sufficiently high level of activity always ensures a liquid-like Newtonian response. The path to this liquified state as a function of increasing activity is however markedly different for values of the target cell shape $p_0$ in the (zero shear, zero activity) solid and liquid phases. In the former, a critical threshold activity level $v_c(p_0)$ is needed to induce liquefaction. In the latter, any level of activity, however small, ensures Newtonian response in  quasistatic shear $\gdot \to 0$. As the shear rate increases, however, the globally coherent effect of shear exceeds that of locally incoherent activity, inducing a resolidification transition via DST.
The shear thickening behavior thus arises from the competition between the accumulation of shear strain due to driving and the dissipation due to cellular activity. On the one hand, the applied shear rate drives the formation of tension networks in the tissue. On the other, the cellular activity acts as a dissipative noisy process that remodels and relaxes the tension network.

We posit that this competition between externally imposed shear and internal cell motility can be characterized via a Peclet number, $Pe = \dot{\gamma}\tau_f$~\cite{Kawasaki2018}, in which $\tau_f$ is the timescale for cell-cell rearrangements due to active motility.
In the regime of high Peclet number, $Pe \gg 1$, $\dot{\gamma} \gg \tau_f^{-1}$, motility is insufficient to affect structural rearrangements caused by the imposed shear. As a result, the mechanical response of the tissue is dominated by the externally applied shear. Because this provides a global driving that acts in a coherent way across the entire tissue, it tends to deform cells away from their target shape, leading to solid-like behaviour.
 
\begin{figure}[tbp]
\includegraphics[width=0.44\textwidth]
{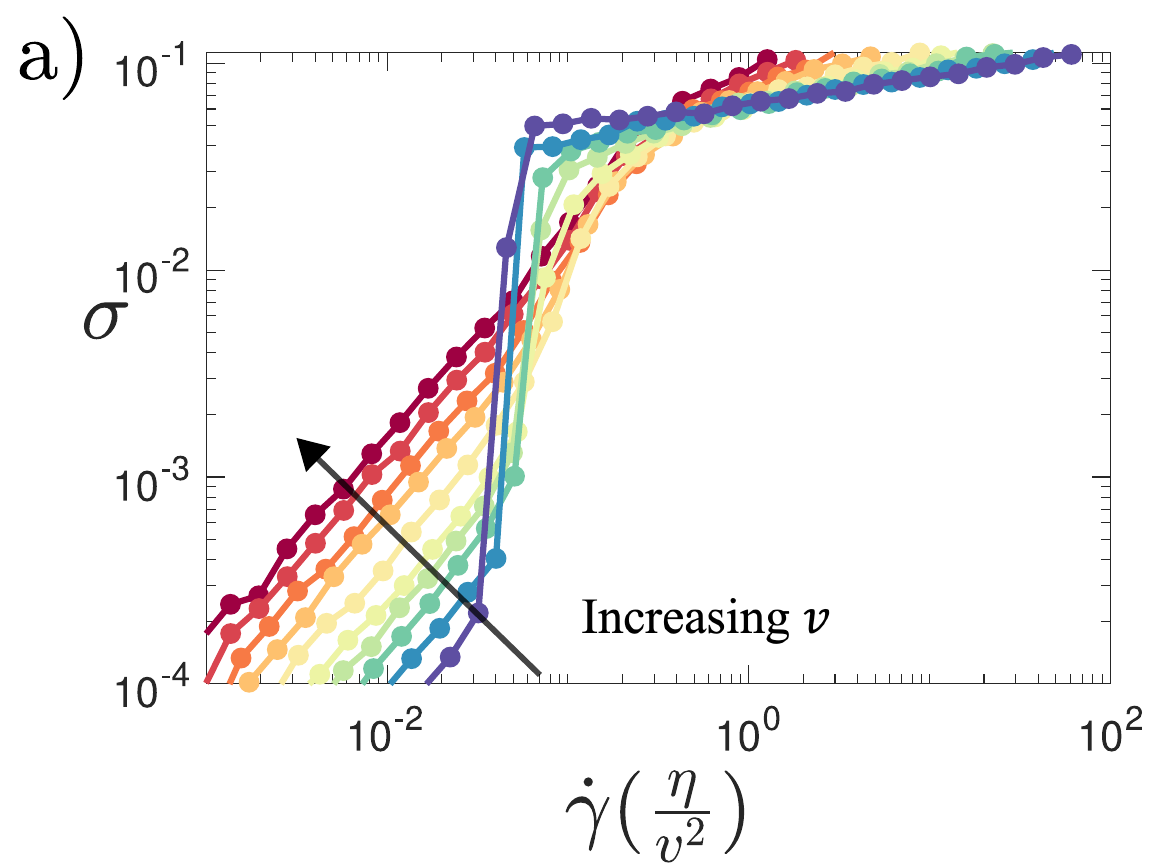}
\includegraphics[width=0.4\textwidth]
{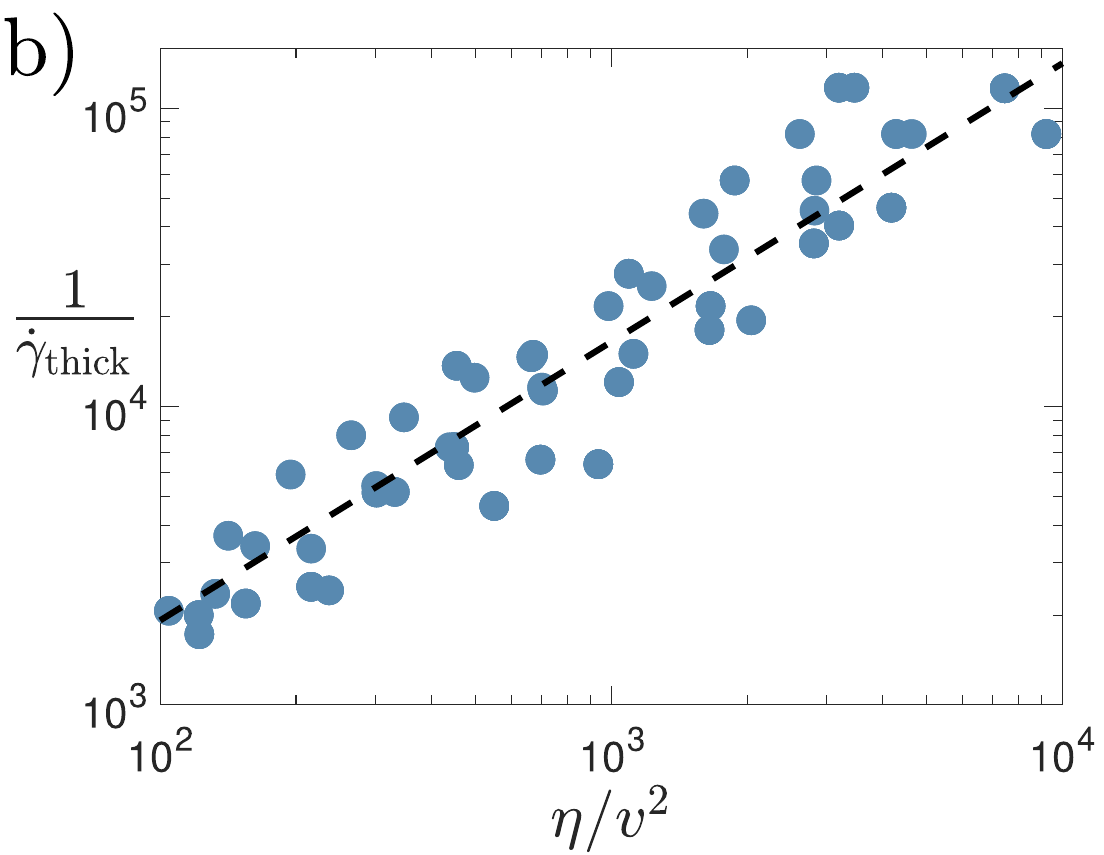}
\caption{Data collapse with Peclet number. {\bf a)} Flow curves for target cell shape $p_0 = 3.90$ and activity $v =0.08, 0.10, 0.12, 0.14, 0.16, 0.20, 0.25, 0.30, 0.35, 0.40$ in curves upwards blue to red at left of figure. Compared with flow curves shown in raw form, as for example in Fig.~\ref{fig:Flow_curves}, the shear rate on the horizontal axis has now been rescaled $\gdot\to\gdot(\eta/v^2)$ to demonstrate scaling collapse with respect to the location of the shear thickening transition. {\bf b)} Inverse shear rate at the shear thickening transition  $1/\gdot_{\rm thick}$ plotted as a function of the scaled viscosity $\eta/v^2$ across the full range of values of cell shape $p_0$ and activity $v$ for which a Newtonian regime and a shear thickening transition are observed in the numerically accessible flow curve. The black dashed line shows a linear scaling $1/\gdot_{\rm thick}\propto \eta/v^2$ as a guide to the eye. }
\label{fig:scaling}
\end{figure}

In contrast, in the regime of low Peclet number, $Pe \ll 1$, $\tau_f^{-1}  \gg \dot{\gamma}$, the mechanics of the tissue is dominated by the active motility. Because this arises internally at the local level of individual cells, lacking any spatial coherence across the tissue, it provides a source of structural rearrangements that tend to counteract the solidifying effect of the globally coherent applied shear just described, resulting in liquid-like response.

We propose that shear thickening occurs at a Peclet number of approximately 1, such that the threshold shear rate for shear thickening is proportional to the inverse of the characteristic time scale $\tau_f$. Deriving an expression for $\tau_f$ is not a simple task. However, we suggest that this timescale for structural rearrangement $\tau_f$ will be proportional to the tissue's Newtonian viscosity $\eta$, defined as the ratio of stress to strain rate in the zero shear rate limit of the flow curve, i.e., at low Peclet number $Pe\ll 1$.

To explore this idea, we show in Fig~\ref{fig:scaling}a) a set of flow curves for a fixed target cell shape $p_0=3.90$, for a range of values of the activity parameter $v$, now with the shear rate on the horizontal axis rescaled according to $\gdot\to\gdot\eta/v^2$.  As can be seen, the location $\gdot_{\rm thick}$ of the shear thickening transition, which is different for different activity values in raw flow curves such as those shown in Fig.~\ref{fig:Flow_curves},  now collapses to a single scaled shear rate, $1/\gdot_{\rm thick} \propto \eta/v^2$. This confirms that the inverse shear rate at which thickening occurs in nonlinear rheology, and so our timescale $\tau_{\rm f}$, is indeed proportional to the tissue's zero shear viscosity $\eta$. This scaling is further investigated in Fig.~\ref{fig:scaling}b), which shows that this relationship $1/\gdot_{\rm thick} \propto \eta/v^2$ is approximately obeyed over the full range of values of $v$ and target cell shape $p_0$ for which both a shear thickening transition and a Newtonian viscosity are indeed seen in the flow curve. 

As just described, this rearrangement timescale $\tau_f$ is important in tissue mechanics because it captures the  timescale for structural and stress relaxation driven by internal activity. Via the above scaling argument, we have demonstrated this quantity $\tau_f$ to be closely  related to the tissue viscosity in the limit of zero shear rate. Importantly, this suggests a possible route to accessing the value of $\tau_f$ experimentally in tissue systems, for example by using magnetically responsive ferrofluid microdroplets to perform quantitative spatiotemporal measurements of mechanical properties in vivo~\cite{Doubrovinski_PNAS,serwane2017vivo}.

In our current model, the focus has been  on the mechanical properties of individual cells and their intercellular interactions, without considering any mechanical role of the cell nucleus. Recent studies have elevated the importance of nuclear compressibility and size as factors that not only govern cell migration and rearrangement but also actively regulate cellular force generation~\cite{Oswald2017,Blauth_review_2021}. In light of these findings, future research should incorporate nuclear mechanics into our existing model, which has already been shown to undergo a density-driven jamming transition~\cite{Grosser2021,Kim_nuclear_jamming_2022}. We anticipate that the model's rheological properties will become increasingly sensitive to nuclear packing density when the size of the nucleus is substantial relative to that of the cell. This could introduce an additional dimension of shear-thickening behavior, similar to the phenomena observed in densely packed particulate systems~\cite{Wyart2014,Prabhu2022,Browne2014}.

In this work, we have considered a bi-disperse distribution of values of cell size. Looking ahead, it would be valuable to furthermore include phenotypic heterogeneity by incorporating distributions of $v,p_0, \kappa_{\rm A}$ and $ \kappa_P$ values, grounded in empirical measurements of single-cell properties. Previous research has demonstrated that such heterogeneity can significantly influence tissue rigidity and fluidity~\cite{Li_tension,fuhs2022rigid}. Consequently, we anticipate that the introduction of mechanical heterogeneity will give rise to intriguing and complex rheological behaviors.

As noted above, DST has been widely observed in dense granular systems with a phenomenology strikingly similar to that reported here for tissues. In each case, a sudden increase in viscosity occurs with increasing shear rate. In each case, this is associated with an intermittent bimodal switching of stress between low and high shear branches, for imposed shear rates in the vicinity of the transition. However, key differences are also notable. In granular systems, DST arises via the development of {\em frictional contacts} between particles, leading to the formation of \emph{compressive} force chains that percolate and bear load across the sample~\cite{Seto2013,Wyart2014}. In contrast, in tissues we predict DST to arise when the globally coherent effects of an applied shear dominate over the local, spatially incoherent effects of cell motility, leading to the formation of \emph{tensile} force chains that percolate and bear load. The possibility of vorticity banding associated with DST in tissues should be investigated in future 3D simulations that allow spatial variations in the vorticity direction, to explore the analogy with vorticity banding associated with DST in granular systems~\cite{chacko2018dynamic}.

 In living tissues, our model  predictions can be immediately tested in the convergent extension of the Drosophila germband epithelium~\cite{butler2009cell, kasza2014spatiotemporal, lye2015mechanical}. This epithelium experiences elongation along the anterior-posterior axis. During this elongation, the germband tissue is subjected to external shearing forces from neighboring structures, such as the ventral furrow, while simultaneously enduring internal forces due to planar polarized contractions driven by myosin II motor activity. This scenario presents a prime example of the interplay between local active forces and global deformations that is central to our theoretical framework. Analyzing the rheological flow curve throughout this process could provide significant insights. With recent technological advances in imaging~\cite{farrell2017segga} and force-measurement techniques~\cite{fischer2014contractile}, such analyses are becoming increasingly attainable.

In summary, our study provides a robust framework for understanding the rheological behavior of biological tissues. Considering that nearly all living tissues are subject to a dynamic interplay between local active forces and global deformations, one of the model's most straightforward yet far-reaching predictions is that this interplay can lead to a competition between the timescales of structural relaxation and of external driving forces. This in turn gives rise to diverse rheological responses, including discontinuous shear thickening. Consequently, we argue that these predictions are broadly applicable to a wide array of biological systems. Furthermore, as mounting evidence increasingly suggests that dense tissues operate near a jamming-unjamming transition, our theoretical contributions offer valuable insights into how tissue mechanics is modulated in proximity to these critical states.

In future work, it would be interesting to extend the concepts explored here to understand whether strongly nonlinear mechanical phenomena such as tissue fracture ~\cite{Harris_PNAS_stretch,bonfanti_fracture_review_2022} and a ductile-to-brittle transition~\cite{Prakash2021} are related to the tissue's ability to shear-thicken.

{\it Code Availability:  ---}The  code used for this paper is available from the  author upon reasonable request.

\section*{Acknowledgements} 
The authors would like to thank Ludovic Berthier for interesting discussions. This project has received funding from the European Research Council (ERC) under the European Union's Horizon 2020 research and innovation programme (grant agreement No. 885146). D.B. acknowledges support from the National Science Foundation (grant no. DMR-2046683), the Alfred P. Sloan Foundation and The Human Frontier Science Program (Ref.-No.: RGP0007/2022) and the NIGMS of the National Institutes of Health under award number R35GM150494.

\appendix

\section{Deriving vertex model forces}

\label{app:forces}

\begin{figure}[h]
    \centering
    \includegraphics[width = 7cm]{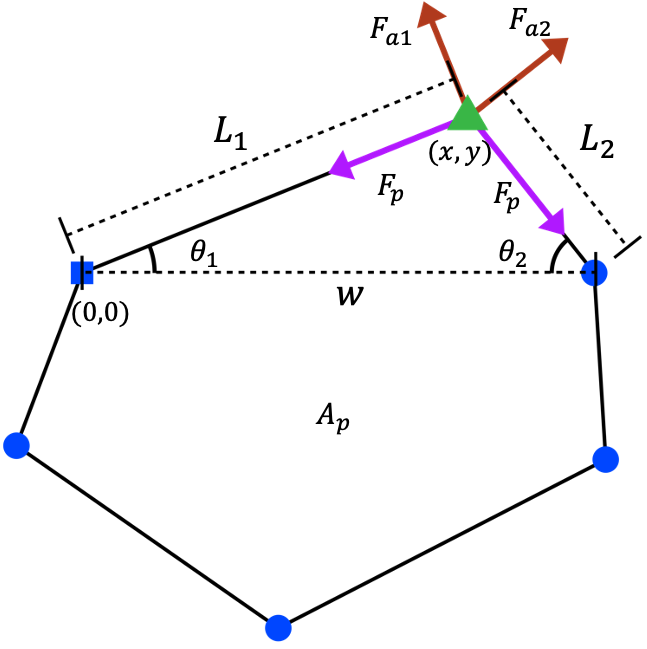}
    \caption{Diagram of forces acting on a single vertex from the two edges of one of the cells that meets that vertex.}
    \label{fig:Diagram}
\end{figure}

We recall from Eqn.~\ref{eqn:energy} in the main text that the elastic energy of a single cell in the vertex model comprises two contributions. The first stems from the deviation of the cell's actual area $A$ from its target value $A_0$. The second stems from the deviation of the cell's actual perimeter $P$ from its target value $P_0$. The resultant force on any given vertex associated with each edge that meets that vertex then likewise comprises separate energy and perimeter contributions, as shown in  Fig.~\ref{fig:Diagram}. (Additional area and perimeter forces also arise associated with the third edge that meets the vertex from neighbouring cells that are not shown.) The area force is shown resolved into components perpendicular to each cell edge. The perimeter force is shown resolved into components parallel to each cell edge. We now derive the magnitude of each of these components.

\subsection{Area Forces}

 The cell sketched in Fig.~\ref{fig:Diagram} can be separated into an upper triangle and a lower pentagon by the bisecting dotted line shown. With the coordinates of the vertex of interest at $(x,y)$ and the origin $(0,0)$ defined to coincide with the vertex to its left, the cell area can be represented as the sum of triangle's area and the pentagon's area:
 \begin{equation}
     A = \frac{1}{2}wy + A_p.
 \end{equation}
In this equation, 
\begin{equation}
\label{eq:w}
    w = L_1\cos(\theta_1) + L_2\cos(\theta_2)
\end{equation}
and $A_p$ is the area of the pentagon. The contribution of this cell to the energy in Eqn.~\ref{eqn:energy} can then be written:
\begin{equation}
    E_a = \frac{1}{2}\kappa_A \left(\frac{1}{2}wy+A_p-A_0\right)^2.
\end{equation}
The $x$ and $y$ components of the area forces can then be calculated by taking the gradient of $E_a$
\begin{equation}  
    \begin{split}
    \nabla E_a 
        =&
        \begin{bmatrix}
        0\\ \frac{1}{2}\kappa_A w(A_0 - A) 
        \end{bmatrix}, \\
        =&
        \begin{bmatrix}
        -F_{a1}\sin(\theta_1) + F_{a2}\sin(\theta_2)\\ F_{a1}\cos(\theta_1) + F_{a2}\cos(\theta_2)
        \end{bmatrix}.
    \end{split}
\end{equation}
On the second line, the $x$ and $y$ component forces are expressed in terms of the components $F_{a1}$ and $F_{a2}$ perpendicular to the cell edges. The $x$ component then yields the equation:
\begin{equation}
0 = -F_{a1}\sin(\theta_1) + F_{a2}\sin(\theta_2),
\label{eq:x_balance}
\end{equation}
and the $y$ component yields the equation:
\begin{equation}
\frac{1}{2}\kappa_A w(A_0 - A)  = F_{a1}\cos(\theta_1) + F_{a2}\cos(\theta_2).
\label{eq:y_balance}
\end{equation}
Recognising that
\begin{equation}
L_1\sin(\theta_1) = L_2\sin(\theta_2)
\end{equation}
then gives, together with Eqn.~\ref{eq:x_balance}, 
\begin{equation}
F_{a2} = F_{a1}\frac{L_2}{L_1}.
\label{eq:x_final}
\end{equation}
Substituting this into Eq.~\ref{eq:y_balance} gives
\begin{equation}
\frac{1}{2}\kappa_A w(A_0 - A)  = F_{a1}\cos(\theta_1) + F_{a1}\frac{L_2}{L_1}\cos(\theta_2).
\label{eq:subbed}
\end{equation}
Rearranging this equation gives
\begin{equation}
F_{a1} = \frac{\frac{1}{2}\kappa_A wL_1(A_0 - A)}{L_1\cos(\theta_1) + L_2\cos(\theta_2)}.
\label{eq:Fa1}
\end{equation}
Substituting the expression for $w$ from Eq.~\ref{eq:w} then gives 
\begin{equation}
    F_{a1} = \frac{1}{2}\kappa_AL_1(A_0-A).
\end{equation}
Substituting this into Eq.~\ref{eq:x_final} likewise gives 
\begin{equation}
    F_{a2} = \frac{1}{2}\kappa_AL_2(A_0-A).
\end{equation}

\subsection{Perimeter Forces}

The perimeter forces can be derived using a similar method. We start by separating the cell perimeter into the contribution from the triangle, $L_1 + L_2$, and the contribution from the pentagon $P_p$ (this being the pentagon's perimeter minus $w$): 
\begin{equation}
    P = L_1 + L_2 + P_p,
    \label{eq:P_expand}
\end{equation}
In this equation,
\begin{equation}
    L_1 = \sqrt{x^2+y^2}
    \label{eq:L_1_def}
\end{equation}
and
\begin{equation}
    L_2 = \sqrt{(w-x)^2+y^2}.
        \label{eq:L_2_def}
\end{equation}
The contribution of this cell to the energy in Eqn.~\ref{eqn:energy} can then be written:
\begin{equation}
    E_p = \frac{1}{2}\kappa_p(\sqrt{x^2+y^2}+\sqrt{(w-x)^2+y^2} + P_p - P_0)^2
\end{equation}
The $x$ and $y$ components of the perimeter forces can then be calculated by taking the
gradient of $E_p$:
  \begin{equation}
   \begin{split}
    \nabla E_p 
        =&
        \begin{bmatrix}
		\kappa_p(\frac{x}{L_1} - \frac{w - x}{L_2})(P - P_0)
		\\ 
		\kappa_p(\frac{y}{L_1} + \frac{y}{L_2})(P - P_0)
  \end{bmatrix}, \\
        =&
         \begin{bmatrix}
        F_P(\cos(\theta_1) - \cos(\theta_2))\\ F_P(\sin(\theta_1) + \sin(\theta_2))
        \end{bmatrix} 
  .
  \end{split}
  \end{equation}
On the first line of this equation, we have substituted expressions for $P, L_1$ and $L_2$ from Eqns.~\ref{eq:P_expand},~\ref{eq:L_1_def} and~\ref{eq:L_2_def}. Further recognising that
\begin{equation}
    \frac{x}{L_1} = \cos(\theta_1),\;\; \frac{w - x}{L_2} = \cos(\theta_2),
    \label{eq:costhet1}
\end{equation}
together with
\begin{equation}
    \frac{y}{L_1} = \sin(\theta_1),\;\;\frac{y}{L_2} = \sin(\theta_2),
    \label{eq:sinthet2}
\end{equation}
gives finally 
\begin{equation}
    F_{P} = \kappa_p(P-P_0).
\end{equation}


%

\end{document}